\documentclass[twocolumn,superscriptaddress]{revtex4-2}
\usepackage{hyperref}
\usepackage{xcolor}
\usepackage{physics}
\usepackage{amsmath}
\usepackage{amssymb}
\usepackage{comment}
\usepackage{algpseudocode}
\usepackage{subfigure}
\usepackage{graphicx}
\usepackage{array,multirow}

\usepackage{qcircuit}\usepackage{xcolor}   
\usepackage{lineno}

\begin{document}
\title{Reducing the Resources Required by ADAPT-VQE Using Coupled Exchange Operators and Improved Subroutines}

\author{Mafalda Ramôa}
\email{mafalda@vt.edu}
\affiliation{Department of Physics, Virginia Tech, Blacksburg, VA, 24061, USA}
\affiliation{Virginia Tech Center for Quantum Information Science and Engineering, Blacksburg, VA 24061, USA}
\affiliation{International Iberian Nanotechnology Laboratory (INL), Portugal}
\affiliation{High-Assurance Software Laboratory (HASLab), Portugal}
\affiliation{Department of Computer Science, University of Minho, Portugal}

\author{Panagiotis G. Anastasiou}
\affiliation{Department of Physics, Virginia Tech, Blacksburg, VA, 24061, USA}
\affiliation{Virginia Tech Center for Quantum Information Science and Engineering, Blacksburg, VA 24061, USA}

\author{Luis Paulo Santos}
\affiliation{International Iberian Nanotechnology Laboratory (INL), Portugal}
\affiliation{High-Assurance Software Laboratory (HASLab), Portugal}
\affiliation{Department of Computer Science, University of Minho, Portugal}

\author{Nicholas J. Mayhall}
\affiliation{Virginia Tech Center for Quantum Information Science and Engineering, Blacksburg, VA 24061, USA}
\affiliation{Department of Chemistry, Virginia Tech, Blacksburg, VA, 24061, USA}

\author{Edwin Barnes}
\affiliation{Department of Physics, Virginia Tech, Blacksburg, VA, 24061, USA}
\affiliation{Virginia Tech Center for Quantum Information Science and Engineering, Blacksburg, VA 24061, USA}

\author{Sophia E. Economou}
\affiliation{Department of Physics, Virginia Tech, Blacksburg, VA, 24061, USA}
\affiliation{Virginia Tech Center for Quantum Information Science and Engineering, Blacksburg, VA 24061, USA}

\begin{abstract}
Adaptive variational quantum algorithms arguably offer the best prospects for quantum advantage in the Noisy Intermediate-Scale Quantum era. Since the inception of the first such algorithm, the Adaptive Derivative-Assembled Problem-Tailored  Variational Quantum Eigensolver (ADAPT-VQE), many improvements have appeared in the literature. We combine the key improvements along with a novel operator pool---which we term Coupled Exchange Operator (CEO) pool---to assess the cost of running state-of-the-art ADAPT-VQE on hardware in terms of measurement counts and circuit depth. We show a dramatic reduction of these quantum computational resources compared to the early versions of the algorithm: CNOT count, CNOT depth and measurement costs are reduced by up to 88\%, 96\% and 99.6\%, respectively, for molecules represented by 12 to 14 qubits (LiH, H$_6$ and BeH$_2$). We also find that our state-of-the-art CEO-ADAPT-VQE outperforms the Unitary Coupled Cluster Singles and Doubles ansatz, the most widely used static VQE ansatz, in all relevant metrics, and offers a five order of magnitude decrease in measurement costs as compared to other static ans\"atze with competitive CNOT counts. \end{abstract}

\maketitle

\section{Introduction}

Quantum computers are expected to be capable of efficiently simulating quantum systems even when this task is intractable for classical computers \cite{feynman}. In the era of noisy intermediate-scale quantum (NISQ) computers, many research efforts have been dedicated to the development of algorithms which are amenable to near-term hardware. One of the leading options for molecular simulations is the variational quantum eigensolver (VQE) proposed in Ref.~\cite{Peruzzo2014}. In this hybrid algorithm, a classical computer is used to minimize a cost function evaluated on a quantum computer. The hope is that the iterative nature of VQE will allow it to employ shallow, NISQ-friendly circuits, in contrast with the long sequences of gates required to execute fully quantum algorithms such as Kitaev's quantum phase estimation \cite{NielsenChuang}. VQE can be employed to find the ground state of many-body systems by taking the energy as the cost function. The expectation value of a Hamiltonian transformed by a many-body unitary is hard to evaluate classically, which suggests that VQE is a good contender for the first demonstration of practical quantum advantage.

In the context of VQE, the \textit{ansatz} is a circuit that applies a parameterized unitary $U(\vec{\theta})$ to a reference state $\ket{\psi_{\mathrm{ref}}}$. Typically, $\ket{\psi_{\mathrm{ref}}}$ is a simple unparameterized state, such as the all-zero state or a classical approximation to the solution, and can be prepared with a constant-depth circuit. $U(\vec{\theta})$ determines, not only the circuit depth and gate count, but also the optimization costs, since it impacts the optimization landscape. The seminal work of Ref.~\cite{Peruzzo2014} proposed the use of the unitary coupled cluster singles and doubles (UCCSD) ansatz. The corresponding unitary is the exponential of a linear combination of single and double fermionic excitations, with the parameter vector $\vec{\theta}$ consisting of the weights of the excitation operators. In contrast, Ref.~\cite{Kandala2017} proposed the hardware-efficient ansatz (HEA), which takes inspiration from device-specific (rather than problem-specific) information to construct the state preparation circuit. The entangling structure of the HEA is based on the connectivity of the device. This ansatz was later shown to suffer from \textit{barren plateaus} (BPs) \cite{McClean2018}, phenomena characterized by the exponential (in the system size) concentration of the cost landscape towards the mean value, calling into question the trainability of such ans\"atze. This problem has since been extensively studied, with results evolving from case-specific proofs of occurrence to a unified theory of BPs \cite{fontana2023,ragone2023unified,diaz2023showcasing}. Several approaches have been proposed to avoid BPs, namely using shallow circuits and local cost functions \cite{Uvarov_2021,Pesah2021}, initializing the ansatz to identity \cite{Grant2019}, and leveraging problem-specific symmetries in the circuit design \cite{Meyer2023}. However, Ref.~\cite{cerezo2023dequantizedvqes} showed that all these strategies collaterally open the door to classical simulation algorithms: The same restrictions on the search space that make variational quantum algorithms provably BP-free can be leveraged to `dequantize' them and dispense with the quantum-classical optimization loop altogether.

One of the few variational quantum algorithms which seem to combine the enticing attributes of being BP-free and not classically simulable is ADAPT-VQE \cite{Grimsley_2019}. While not rigorously proved, the absence of BPs is suggested by both theoretical arguments and empirical evidence \cite{Grimsley2023,larocca2024,cerezo2023dequantizedvqes}. The special feature of this algorithm is that the ansatz is constructed dynamically, by iteratively appending to a reference circuit parameterized unitaries generated by elements selected from an operator pool. The screening of generators is based on their energy derivatives (usually referred to simply as `gradients'), such that at each step the choice of unitary depends on the variational state as well as on the molecular Hamiltonian. This problem- and system-tailored approach leads to remarkable improvements in circuit efficiency, accuracy, and trainability with respect to fixed-structure ans\"atze \cite{Grimsley_2019,Grimsley2023}. These advantages have instigated the community to further understand and develop the algorithm. Active research topics include reducing measurement costs \cite{anastasiou2023, Liu2021, Majland2023, Nykanen2022, nakagawa2023, ramôa2024}, seeking minimal complete pools (i.e., pools of minimal size that enable convergence) \cite{shkolnikov2021}, decreasing circuit depth \cite{anastasiou2022,long_2023, Feniou2023,Fitzpatrick_2024}, constructing more circuit-efficient pools \cite{Tang_2021,Yordanov_2021}, studying the effect of noise \cite{Dalton2022}, bypassing the optimization process \cite{Gomes2021}, and generalizing the algorithm to excited states \cite{yordanov_2021excited,Zhang2021,nykänen2024deltaadaptvqe} or to other problems \cite{Zhu2020,warren2022,Romero2022,Yao2021,linteau2023adaptive,nykänen2023,Yoshikura2023,Majland2024}. Recently, a variant of ADAPT-VQE termed Scalable Circuits (SC)-ADAPT-VQE was used to prepare a 100-qubit vacuum state of the Schwinger model on a superconducting-qubit quantum computer \cite{farrell_2024}.

Despite this intense research following up on ADAPT-VQE, and the state of the art improving dramatically since the first paper \cite{Grimsley_2019}, it is still not understood how close we are to a demonstration on near-term hardware that can rival classical simulation. In fact, there is significant skepticism from the community that pre-fault tolerant quantum algorithms can ever demonstrate quantum advantage. One of the often-cited concerns is the large number of measurements associated with VQEs.

Here, we take a key step in addressing this question by introducing an improved operator pool, which we term coupled exchange operator (CEO) pool, and merging this with recent progress in decreasing the measurement costs and improving the hardware efficiency of the adaptive ansatz construction. We show that this novel variant of ADAPT-VQE, labeled CEO-ADAPT-VQE*, combines frugal measurement costs with shallow ans\"atze. We use this to gauge the evolution of ADAPT-VQE since its inception. Our simulations for a range of molecules show that our present version outperforms all previous ones, and the gate counts are reduced dramatically as compared to the original ADAPT-VQE. Further, the measurement costs incurred by our algorithm are five orders of magnitude lower than those incurred by static ans\"atze with comparable CNOT counts. These improvements bring us closer to the goal of demonstrating practical quantum advantage.

\begin{table*}[htbp]
  \centering
\begin{center}
\begin{tabular}{|c|c|c|c|c|}
\cline{3-5}
\multicolumn{2}{c|}{} & {LiH at 3\AA{}} & {H$_6$ at 1.5\AA{}} & {BeH$_2$ at 2\AA{}} \\
\hline
\multirow{ 3}{*}{GSD-ADAPT-VQE} & CNOT Count & 392 & 6896 & 2192\\
\cline{2-5}
& CNOT Depth & 384 & 6586 & 1909 \\
\cline{2-5}
& Measurement Costs & 50468 & 654899 & 520225 \\
\hline  
\multirow{ 3}{*}{CEO-ADAPT-VQE*}& CNOT Count (reduced to) & 107 (27\%) & 812 (12\%) & 288 (13\%) \\
\cline{2-5}
& CNOT Depth (reduced to) & 30 (8\%) & 282 (4\%) & 95 (5\%) \\
\cline{2-5}
& Measurement Costs (reduced to) & 560 (1\%) & 10857 (2\%) & 2197 (0.4\%) \\
\hline
\end{tabular}
\end{center}
  \caption{Minimal resource requirements of GSD-ADAPT-VQE and CEO-ADAPT-VQE* to reach chemical accuracy (defined as an error below 1kcal/mol). The algorithms and costs are defined in the main text (see, e.g., Fig.~\ref{fig:combining_proposals_circuit}).}
  \label{tab:resource_analysis}
\end{table*}

Table~\ref{tab:resource_analysis} showcases the evolution of ADAPT-VQE to date by comparing our novel algorithm with GSD-ADAPT-VQE for LiH (12 qubits), H$_6$ (12 qubits) and BeH$_2$ (14 qubits). As proposed in the original formulation of ADAPT-VQE, the latter uses a fermionic pool, consisting of generalized single and double (GSD) excitations. The table presents the CNOT count, CNOT depth and measurement costs (estimated as the total number of noiseless energy evaluations, which constitutes a lower bound) of the algorithms at the first iteration where they reach chemical accuracy. We observe that CEO-ADAPT-VQE* reduces these costs dramatically---to 12-27\%, 4-8\% and 0.4-2\%, respectively, as compared to the original ADAPT-VQE algorithm. These reductions bring us closer to the goal of demonstrating practical quantum advantage.

The paper is structured as follows. We start the results section, Sec.~\ref{s:results}, with Sec.~\ref{ss:background}, which reviews the VQE and ADAPT-VQE algorithms. Section \ref{ss:CEO} inspects the structure of qubit excitations (Sec.~\ref{sss:CEO_motivation}) to motivate the definition of the coupled exchange operators (Secs.~\ref{sss:mvp_ceo} and \ref{sss:ovp_ceo}), which is used to define the CEO-ADAPT-VQE algorithm in Sec.~\ref{sss:CEO-ADAPT}. Section \ref{ss:classical_sims} contains the results of numerical simulations comparing the newly proposed algorithm with previous variants. In Sec.~\ref{sss:convergence_plots}, we compare the iteration, parameter, and CNOT counts of CEO-, QEB- and Qubit-ADAPT-VQE for various molecules at different geometries. Section~\ref{sss:bond_diss} offers a comparison of CEO-ADAPT-VQE with UCCSD-VQE throughout the bond dissociation curves of the same molecules. In Sec.~\ref{sss:adapt_evolution}, we discuss enhancing CEO-ADAPT-VQE with other improvements proposed in the literature. To showcase the reduction in the resource requirements of ADAPT-VQE from the time it was first proposed, we compare this enhanced version with a fermionic variant of the algorithm as proposed in the original work of Ref.~\cite{Grimsley_2019}. Finally, in Sec.~\ref{sss:adapt_vs_fixed} we compare our adaptive algorithm with the leading fixed-structure ans\"atze. Section ~\ref{s:discussion} contains concluding remarks.

\section{Results}
\label{s:results}

\subsection{Background}
\label{ss:background}

This section provides the theoretical background that is fundamental to understand the paper. We discuss the variational quantum eigensolver in Sec.\ref{sss:VQE} and its adaptive version, ADAPT-VQE, in Sec.~\ref{sss:adapt}. Readers familiar with VQE and ADAPT-VQE could skip this section.

\subsubsection{VQE}
\label{sss:VQE}

The variational quantum eigensolver 
 \cite{Peruzzo2014} is a hybrid quantum-classical algorithm designed to find eigenstates and eigenvalues of physical systems. We are interested in its application to the electronic structure problem. More precisely, we seek solutions of the time-independent Schrödinger equation
\begin{equation}
\hat{\mathcal{H}}\ket{\psi}=E\ket{\psi}\,,
\label{eq:schrodinger}
\end{equation}
where $\hat{\mathcal{H}}$ is the electronic Hamiltonian arising from the Born–Oppenheimer approximation. The solutions to this equation are stationary states $\ket{\psi}$ with corresponding energy $E$. We will focus on the task of finding the ground state $\ket{\psi_0}$ and the ground energy $E_0$. We refer to Refs.~\cite{Higgott_2019,Colless_2018,yordanov_2021excited,Asthana_ChemSci2023} for approaches targeting excited states.

The variational principle of quantum mechanics states that
\begin{equation}
    \bra{\psi}\hat{\mathcal{H}}\ket{\psi}\geq E_0
    \label{ineq:variational_principle}
\end{equation}
for any normalized state $\ket{\psi}$. Variational methods for the ground state problem are based on the variational principle: They define a suitable parameterized wave function $\ket{\psi(\vec{\theta})}$ and minimize the energy by adjusting the vector $\vec{\theta}$. The variational form $\ket{\psi(\vec{\theta})}$, also called the ansatz, determines the search space. Evidently, a variational method can only find the ground state if it is contained within the ansatz.

There are many classical ansätze for the electronic structure problem. Because the memory required to store a generic electronic wave function on a classical computer grows exponentially with the number of orbitals included in the basis set, exact classical algorithms quickly become intractable. On the other hand, approximate alternatives often yield insufficient accuracy. We refer to Ref.~\cite{SzaboOstlund} for an overview of this topic.

A natural alternative is to prepare the variational state on a quantum computer, where the memory requirements scale only linearly with the number of orbitals. This is the motivation behind VQE \cite{Peruzzo2014}. In this algorithm, the role of the quantum computer is to prepare parameterized fermionic states and measure their energy. A classical computer is employed to minimize the energy by tuning the parameters. VQE is meant to be a NISQ-friendly alternative to fully quantum approaches, since the hybrid and iterative procedure leads to shallower circuits.

A physically motivated choice of ansatz for VQE is the UCCSD ansatz \cite{Peruzzo2014,romero2018}, inspired by classical variational algorithms and their limitations. The corresponding variational form is defined as
\begin{equation}
    \ket{\mathrm{UCCSD}}={e^{(T_1+T_2)-(T_1^\dagger+T_2^\dagger)}}\ket{\mathrm{HF}},
    \label{def:UCCSD_operator}
\end{equation}
where $\ket{\mathrm{HF}}$ is the Hartree-Fock reference state (obtained from self-consistent mean-field calculations on a classical computer), and $T_1$ and $T_2$ are operators which generate single and double excitations:
\begin{equation}
\begin{split}
T_1=\sum_{i>a}&t_i^aa_a^\dagger a_i,\\
T_2=\sum_{i>j,a>b}&t_{ij}^{ab}a_a^\dagger a_b^\dagger a_i a_j\,,\\
\end{split}
\label{def:fermionic_excitations}
\end{equation}
where $i$, $j$ ($a$, $b$) correspond to orbitals that are occupied (unoccupied) in the reference state. The $a_i$/$a_i^\dagger$ are fermionic ladder operators which respectively remove/add a fermion from/to the $i$th spin-orbital. We can also use these operators to write the fermionic Hamiltonian as
\begin{equation}
    \mathcal{\hat{H}} = \sum_{p,q}^N h_{p,q}a_p^\dagger a_q + \sum_{p,q,r,s}^N h_{p,q,r,s} a_p^\dagger a_q^\dagger a_r a_s\,,
    \label{eq:hamiltonian}
\end{equation}
where $h_{p, q}$ ($h_{p, q, r, s}$) are one- (two-) electron integrals, and $p$, $q$, $r$, $s$ run over all $N$ spin-orbitals. In order to carry out the VQE algorithm, we need to be able to prepare the variational state in Eq.~\eqref{def:UCCSD_operator} and measure the fermionic Hamiltonian in Eq.~\eqref{eq:hamiltonian}. These two tasks require a mapping from fermionic operators to qubit operators. The Jordan-Wigner mapping \cite{JordanWigner},
\begin{align}
\begin{split}
a_i^\dagger\rightarrow \frac{1}{2}\prod_{k=1}^{i-1}Z_k\cdot(X_i-iY_i),\\
\quad
a_i\rightarrow \frac{1}{2}\prod_{k=1}^{i-1}Z_k\cdot(X_i+iY_i),
\end{split}
\label{def:jw_transform_paulis}
\end{align}
is a convenient choice. $Z_k$, $X_i$ and $Y_i$ are Pauli operators acting on the qubits labeled by the respective indices. 

After applying the transformation, we can implement the UCCSD state preparation circuit with standard tools from quantum simulation \cite{NielsenChuang} and measure the Hamiltonian in the quantum computer via sampling \cite{McClean2016}. The resulting circuit is a system-agnostic composition of unitaries generated by single and double excitations.

\subsubsection{ADAPT-VQE}
\label{sss:adapt}

While the UCCSD ansatz is an interesting option for VQE, it has a few shortcomings: (i) it fails to reach chemical accuracy (defined as an error below 1kcal/mol) for some systems, (ii) it includes all single and double excitations, despite it being expected that only a system-dependent subset is relevant, and (iii) it is rendered ambiguous by Trotterization \cite{Grimsley2020}.

ADAPT-VQE \cite{Grimsley_2019} was proposed to tackle these issues. In this algorithm, the ansatz is constructed dynamically, in a way dictated by the system under study. This was shown to result in lower errors, shallower circuits and lower parameter counts than UCCSD-VQE. One of the key elements of this algorithm is an \textit{operator pool} $\{\hat{A}_k\}_K$, a set of anti-Hermitian operators which must be defined in advance. We summarize the ADAPT-VQE protocol in five steps:

\begin{enumerate}
    \item Initialize the variational state to a classically efficient reference state (e.g., the Hartree-Fock ground state: $\ket{\psi^{(0)}}=\ket{\mathrm{HF}}$) and the iteration counter to 1 ($n\gets1$).
    \item For each operator $\hat{A}_k$ in the pool, evaluate the derivative of the energy with respect to the variational parameter $\theta_k$ when the unitary $e^{\theta_k \hat{A}_k}$ is appended to the current ansatz with $\theta_k=0$. If the norm of the vector formed by these gradients is under a pre-defined convergence threshold $\epsilon$, terminate. 
    \label{step:meas_gradients}
    \item Select the generator $\hat{A}_n$ associated with the highest magnitude energy derivative and append the corresponding unitary to the ansatz: $\ket{\psi^{(n)}} = e^{\theta_n\hat{A}_n}\ket{\psi^{(n-1)}}$, where $\theta_n$ is a new variational parameter.
    \label{step:selection}
    \item Obtain the new parameter vector $\vec{\theta}^{(n)}$ and energy $E^{(n)}$ from a VQE optimization over all parameters, initialized at $\{\vec{\theta}^{(n-1)},0\}$. 
    \item Increment the iteration counter ($n\gets n+1$) and go to step \ref{step:meas_gradients}.
\end{enumerate}

We note that some variants of the algorithm may not strictly abide by this generic workflow. For example, alternative termination criteria can be used in step \ref{step:meas_gradients}, such as a threshold on the energy change or maximum gradient magnitude. Step \ref{step:selection} can also employ alternative selection criteria. The criterion in Ref.~\cite{Yordanov_2021} is based on the energy changes obtained from optimizing multiple ans\"atze, each constructed by appending a unitary whose gradient magnitude ranks among the highest. This amounts to repeating step \ref{step:selection} for multiple candidates in an attempt to maximize the decrease in the energy per iteration. However, this incurs a significant increase in measurement costs without consistently improving the performance \cite{yordanov2020sup}. To minimize costs and allow for a fair comparison between variants, we will exclusively consider the criterion proposed in the original work, Ref.~\cite{Grimsley_2019}.

The derivatives in step \ref{step:meas_gradients} are often referred to as \textit{gradients} for simplicity. Similarly, we may relax the nomenclature and refer to the gradient of the unitary generated by $\hat{A}_k$ as `the gradient of $\hat{A}_k$'. This creates no ambiguity in our context, as there is no circumstance in which we consider the gradients of the generators.

The gradients can be measured in a quantum computer using the commutator formula
\begin{align}
\begin{split}
&\frac{\partial \bra{\psi^{(n-1)}}e^{-\theta_k\hat{A}_k}\mathcal{\hat{H}}e^{\theta_k\hat{A}_k}\ket{\psi^{(n-1)}}} {\partial\theta_k}\Bigr|_{\theta_k = 0} \\
&=\bra{\psi^{(n-1)}}[\mathcal{\hat{H}},\hat{A}_k]\ket{\psi^{(n-1)}}.
\end{split}
\label{eq:adapt_commutator}
\end{align}
Since the value of this commutator must be obtained for each pool element, the choice of pool impacts the measurement costs of the algorithm. It also impacts the circuit efficiency and solution quality, since all constituents of the final ansatz will be parameterized unitaries generated by the pool operators. We will focus on the two leading pools (in terms of hardware-efficiency): The qubit excitation (QE) pool \cite{Yordanov_2021} and the qubit pool \cite{Tang_2021}. 

The QE pool is comprised of Jordan-Wigner-transformed generalized single and double fermionic excitations from which the anticommutation string ($\prod_{k=1}^{i-1}Z_k$ in Eq.~\eqref{def:jw_transform_paulis}) is removed. These are essentially the summands in Eq.~\eqref{def:fermionic_excitations}, except here we consider \textit{generalized} excitations instead of restricting the source/target orbitals to be occupied/unoccupied in the reference state. The operators in this pool preserve particle number and $S_z$, but they do not faithfully represent the fermionic anticommutation relations. QEs obtained from single excitations act on two qubits and consist of linear combinations of two Pauli strings, while those obtained from double excitations act on four qubits and consist of eight Pauli strings. The unitaries generated by these operators (often called `QE evolutions') can be implemented using circuits with 2 and 13 CNOTs, respectively \cite{yordanov2020circuits,Yordanov_2021,Wang_2021}. 

Qubit pools \cite{Tang_2021} consist of individual Pauli strings. In general, they do not conserve particle number or $S_z$, nor do they respect anticommutation. The corresponding evolutions are straightforwardly implemented using ladder-of-CNOTs circuits \cite{NielsenChuang}. We consider the qubit pool formed from all individual Pauli strings appearing in the QE pool. Each one acts on two or four qubits, depending on whether it originated from a single or double excitation. The unitaries generated by them can be implemented using circuits with 2 or 6 CNOTs, respectively.

The two pools described above define the Qubit Excitation Based (QEB)-ADAPT-VQE \cite{Tang_2021} and the Qubit-ADAPT-VQE \cite{Yordanov_2021} algorithms.

\subsection{Coupled Exchange Operator (CEO) - ADAPT-VQE}
\label{ss:CEO}

The choice of operator pool dictates the structure of the unitaries that will appear in the ansatz. Drawing inspiration from unitary coupled cluster theory, the first version of ADAPT-VQE \cite{Grimsley_2019} used a pool consisting of anti-Hermitian sums of fermionic excitations mapped into qubit operators via the Jordan-Wigner transform \cite{JordanWigner}. However, the corresponding evolutions (exponentials generated by these operators) are not convenient to implement with a quantum circuit. In order to preserve particle number and total $Z$ spin projection ($S_Z$), they must correspond to a linear combination of several Pauli strings; and due to the nonlocality inherent to the antisymmetry of the fermionic many-body wave function, the strings must grow linearly with the number of orbitals in the basis set. This means that the number of entangling gates required to implement the unitary generated by each pool operator will grow linearly (on average) with the number of qubits. This was improved in Ref.~\cite{Tang_2021}, which proposed Qubit-ADAPT-VQE, a variant of the algorithm where the pool consists of individual Pauli strings. They showed that a pool of operators which do not respect fermionic anticommutation nor preserve particle number and $S_Z$ can produce circuits with a significantly lower CNOT count than the original fermionic pool, at the expense of a higher number of variational parameters and measurements. Circuit efficiency was further improved in Ref.~\cite{Yordanov_2021}, which proposed a pool of \textit{qubit excitations} (QEs), operators which preserve particle number and $S_Z$ symmetries despite not respecting anticommutation. The symmetry-preserving structure of QEs can be leveraged to create CNOT-efficient circuits for the corresponding evolutions \cite{yordanov2020circuits}, and the resulting qubit-excitation-based (QEB)-ADAPT-VQE algorithm stands as the most hardware-efficient variant as of today.

In this section, we propose CEO-ADAPT-VQE, an adaptive VQE based on coupled exchange operators (CEOs). These operators consist of linear combinations of QEs acting on the same set of spin-orbitals, which may share one variational parameter (OVP-CEOs) or be independently parameterized (MVP-CEOs). These operators are capable of simultaneously realizing exchanges between multiple pairs of Slater determinants, in contrast with QEs, which exchange exactly one pair of determinants each. We show that evolutions of MVP-CEOs consisting of up to three QEs can be implemented by circuits with the same CNOT count as the evolution of each individual excitation, while OVP-CEO evolutions can be implemented by circuits with roughly 30\% \textit{fewer} CNOTs. The CNOT counts are the key cost to consider when dealing with NISQ devices, as they are by far associated with the highest error rates. We provide explicit constructions for all circuits and numerically simulate CEO-ADAPT-VQE for multiple molecules. The results show that this algorithm reduces the CNOT count with respect to both QEB- and Qubit-ADAPT-VQE, with a decrease of roughly 50\% and 65\% (respectively) for the most difficult systems. The difference is expected to increase with the system size. This decrease does not entail any collateral increase in measurement costs. In fact, with respect to the other variants, our algorithm maintains or decreases the total number of gradient measurement rounds as well as of variational parameters.

We will motivate the introduction of these operators with a detailed analysis of QEs (Sec.~\ref{sss:CEO_motivation}). Secs.~\ref{sss:mvp_ceo} and \ref{sss:ovp_ceo} introduce two different classes of CEOs. Finally, the complete CEO-ADAPT-VQE algorithm is proposed in Sec.\ref{sss:CEO-ADAPT}.

\subsubsection{Motivation}
\label{sss:CEO_motivation}

We begin by delving into the structure of double QEs \cite{Yordanov_2021} and the corresponding circuit implementation.

Suppose we have a set of four qubits, two of which correspond (under the Jordan-Wigner mapping) to $\alpha$-type spin-orbitals and two of which correspond to $\beta$-type ones. Accordingly, we label them $\alpha_1$, $\alpha_2$, $\beta_1$, $\beta_2$, where the numeric labels within the same spin-orbital type can be chosen arbitrarily. There will be two unique double QEs acting on these spin-orbitals:
\begin{equation}
T_{\alpha_1\beta_1\rightarrow\alpha_2\beta_2}^{(QE)} = Q^\dagger_{\alpha_2}Q^\dagger_{\beta_2}Q_{\alpha_1}Q_{\beta_1} - Q^\dagger_{\beta_1}Q^\dagger_{\alpha_1}Q_{\beta_2}Q_{\alpha_2},
    \label{eq:qe1}
\end{equation}
\begin{equation}
T_{\alpha_2\beta_1\rightarrow\alpha_1\beta_2}^{(QE)} = Q^\dagger_{\alpha_1}Q^\dagger_{\beta_2}Q_{\alpha_2}Q_{\beta_1} - Q^\dagger_{\beta_1}Q^\dagger_{\alpha_2}Q_{\beta_2}Q_{\alpha_1},
    \label{eq:qe2}
\end{equation}
where 
\begin{align}
\begin{split}
&Q_i^\dagger = \frac{1}{2}(X_i-iY_i),\\
&Q_i = \frac{1}{2}(X_i+iY_i),
    \label{def:qubit_ladder_operators}
\end{split}
\end{align}
are the qubit creation and annihilation operators. They are equivalent to the fermionic creation and annihilation operators after removal of the Jordan-Wigner anticommutation string (see Eq.~\eqref{def:jw_transform_paulis}).

Under the Jordan-Wigner transform, the state of each qubit represents the occupation number of a spin-orbital. Slater determinants are represented by computational basis states whose particle number is given by the Hamming weight of the corresponding bit string, and whose spin quantum number is given by $\frac{1}{2}(N_\alpha-N_\beta)$, where $N_\alpha$ and $N_\beta$ are the total occupation numbers of all qubits representing $\alpha$ and $\beta$ spin-orbitals, respectively. It is easy to see that the operators in Eqs.~\eqref{eq:qe1} and \eqref{eq:qe2} exchange two determinants in such a way that these quantities are preserved. 

Operators $T_{\alpha_2\beta_2\rightarrow\alpha_1\beta_1}^{(QE)}$,  $T_{\alpha_1\beta_2\rightarrow\alpha_2\beta_1}^{(QE)}$ are also valid QEs; however, they differ from the operators in Eqs.~\eqref{eq:qe1}, \eqref{eq:qe2} (respectively) only by a minus sign. The sign reflects which electronic transition we label as an excitation and which we label as a de-excitation. As this labeling becomes irrelevant once the operator is multiplied by a variational parameter, these operators are redundant with the above, and we can freely choose either option for each case. As for the two non-redundant QEs, the difference between them is that the operator in Eq.~\eqref{eq:qe1} exchanges $\ket{0}_{\alpha_2}\ket{1}_{\alpha_1}\ket{0}_{\beta_2}\ket{1}_{\beta_1}$ with $\ket{1}_{\alpha_2}\ket{0}_{\alpha_1}\ket{1}_{\beta_2}\ket{0}_{\beta_1}$, while the operator in Eq.~\eqref{eq:qe2} exchanges $\ket{1}_{\alpha_2}\ket{0}_{\alpha_1}\ket{0}_{\beta_2}\ket{1}_{\beta_1}$ with $\ket{0}_{\alpha_2}\ket{1}_{\alpha_1}\ket{1}_{\beta_2}\ket{0}_{\beta_1}$. All other Slater determinants are quenched.

Henceforth we ignore all qubits on which the operator action is trivial, and assume that the four relevant qubits are ordered as $\alpha_2$, $\alpha_1$, $\beta_2$, $\beta_1$ from the most to the least significant. We can thus omit the indices; e.g., $\ket{0}_{\alpha_2}\ket{1}_{\alpha_1}\ket{0}_{\beta_2}\ket{1}_{\beta_1}$ and $X_{\alpha_2}Y_{\alpha_1}X_{\beta_2}X_{\beta_1}$ are represented simply as $\ket{0101}$ and $XYXX$ using little endian ordering. This choice is merely for the sake of clarity and incurs no loss of generality.

In terms of Pauli strings, the operators can be written as
\begin{align}
\begin{split}
T_{\alpha_1\beta_1\rightarrow\alpha_2\beta_2}^{(QE)} =
&\frac{i}{8}( XXXY - XXYX + XYXX + XYYY \\
&- YXXX - YXYY + YYXY - YYYX ) \\
=&\frac{i}{8}XXXY(1 - IIZZ + IZIZ - IZZI \\
& - ZIIZ + ZIZI - ZZII + ZZZZ),
\end{split}
\label{eq:qe1_paulis}
\end{align}
\begin{align}
\begin{split}
T_{\alpha_2\beta_1\rightarrow\alpha_1\beta_2}^{(QE)} =
&\frac{i}{8}( XXXY - XXYX - XYXX - XYYY \\
&+ YXXX + YXYY + YYXY - YYYX ) \\
=&\frac{i}{8}XXXY(1 - IIZZ - IZIZ + IZZI \\
& + ZIIZ - ZIZI - ZZII + ZZZZ ).
\end{split}
\label{eq:qe2_paulis}
\end{align}
We factored out the $XXXY$ Pauli string to  emphasize the action of each of the operators. We note that the choice of which Pauli string to factor out is irrelevant, as any of the eight would result in the same expression in brackets (with a $-1$ multiplicative factor in half of the cases). It is easy to see that the last bracketed expression in Eq.~\eqref{eq:qe1_paulis} acts on $\ket{0101}$ and $\ket{1010}$ as
\begin{align}
\begin{split}
\ket{0101}&\rightarrow +8\ket{0101},\\
\ket{1010}&\rightarrow +8\ket{1010},
\end{split}
\label{eq:qe1_exchanges}
\end{align}
while quenching all other computational basis states. Similarly, the last bracketed expression in Eq.~\eqref{eq:qe2_paulis} quenches all computational basis states except $\ket{1001}$ and $\ket{0110}$, on which it acts as
\begin{align}
\begin{split}
\ket{1001}&\rightarrow +8\ket{1001},\\
\ket{0110}&\rightarrow +8\ket{0110}.
\end{split}
\label{eq:qe2_exchanges}
\end{align}
On the other hand, $iXXXY$ applies a bit flip (in the computational basis) to each of the qubits it acts on, changing the occupation number of the corresponding spin-orbitals. Additionally, it imparts a phase $\pm 1$ which depends on the value of the 0th (rightmost) qubit. This phase is a reflection of our choice in labeling excitations / de-excitations. 

The operators that constitute the ansatz are parameterized and exponentiated versions of qubit excitations:
\begin{equation}
U_{\alpha_1\beta_1\rightarrow\alpha_2\beta_2}^{(QE)}=e^{\theta T_{\alpha_1\beta_1\rightarrow\alpha_2\beta_2}^{(QE)}},
\label{eq:qe1_expon}
\end{equation}
\begin{equation}
U_{\alpha_2\beta_1\rightarrow\alpha_1\beta_2}^{(QE)}=e^{\theta T_{\alpha_2\beta_1\rightarrow\alpha_1\beta_2}^{(QE)}}.
\label{eq:qe2_expon}
\end{equation}
The unitary in Eq.~\eqref{eq:qe1_expon} will rotate $\ket{0101}$, $\ket{1010}$ as 
\begin{align}
\begin{split}
\ket{0101}&\rightarrow \cos{\theta}\ket{0101}+\sin{\theta}\ket{1010},\\
\ket{1010}&\rightarrow \cos{\theta}\ket{1010}-\sin{\theta}\ket{0101},
\end{split}
\label{eq:qe1_expon_exchanges}
\end{align}
while the one in Eq.~\eqref{eq:qe2_expon} will rotate $\ket{1001}$, $\ket{0110}$ as 
\begin{align}
\begin{split}
\ket{1001}&\rightarrow \cos{\theta}\ket{1001}+\sin{\theta}\ket{0110},\\
\ket{0110}&\rightarrow \cos{\theta}\ket{0110}-\sin{\theta}\ket{1001}.
\end{split}
\label{eq:qe2_expon_exchanges}
\end{align}
All other Slater determinants are left unchanged.

An important question is how to implement the unitaries in Eqs.~\eqref{eq:qe1_expon}, \eqref{eq:qe2_expon} as quantum circuits. Since all the Pauli strings in the exponent commute, no Trotterization is required, and we can implement the exponentials of the eight strings in sequence using eight pairs of three-step CNOT ladders \cite{NielsenChuang}. Naively, this would require 48 CNOTs. However, if the CNOTs between rotations are instead implemented such that they all share the same target, and we organize the Pauli strings such that two consecutive ones differ on two qubits, the CNOT count can be reduced to 13 \cite{Nam_2019,Wang_2021}.

Another alternative, proposed in Ref.~\cite{yordanov2020circuits}, is to leverage the fact that the operators are simply conditional rotations. More precisely, the rotations are applied to a computational basis state or not depending on the parities of the states of some subsets of qubits. For example, the operator in Eq.~\eqref{eq:qe1_expon} applies a rotation to a computational basis state $\ket{x_3x_2x_1x_0}$ if and only if $\overline{x_0 \oplus x_2}\land\overline{x_1  \oplus x_3}\land (x_0\oplus x_1$). The first (second) term guarantees that the occupation number of orbitals $\alpha_1$ and $\beta_1$ ($\alpha_2$ and $\beta_2$) is the same. The last term guarantees that if the former are occupied, the latter are unoccupied or vice-versa. Since CNOTs act as reversible XOR gates, it becomes evident that this operator can be implemented as in Fig.~\ref{cir:qe1_implicit}. Similarly, the operator in Eq.~\eqref{eq:qe2_expon} can be implemented by the circuit in Fig.~\ref{cir:qe2_implicit}.

\begin{figure}[htbp]

    \centerline{
    \Qcircuit @C=1em @R=.7em {
    & \ctrl{2} & \qw & \ctrl{1} & \gate{R_y(-2\theta)} & \ctrl{1} & \qw & \ctrl{2} & \qw \\
    & \qw & \ctrl{2} & \targ & \ctrl{-1}& \targ  & \ctrl{2} & \qw & \qw  \\
    & \targ & \qw & \qw & \ctrlo{-1} & \qw & \qw & \targ  & \qw &\\
    & \qw & \targ & \qw & \ctrlo{-1}& \qw  & \targ & \qw & \qw
    }}
    
    \caption{Circuit implementation of the QE evolution $U_{\alpha_1\beta_1\rightarrow\alpha_2\beta_2}$. The circuit for $U_{\alpha_2\beta_2\rightarrow\alpha_1\beta_1}$ is identical, but with the rotation angle flipped. While the sign is necessary for the circuit to correspond exactly to this operator, it becomes irrelevant when $\theta$ is optimized variationally.}

\label{cir:qe1_implicit}
\end{figure}

\begin{figure}[htbp]

    \centerline{
    \Qcircuit @C=1em @R=.7em {
    & \ctrl{3} & \qw & \ctrl{1} & \gate{R_y(-2\theta)} & \ctrl{1} & \qw & \ctrl{3} & \qw \\
    & \qw & \ctrl{1} & \targ & \ctrl{-1}& \targ  & \ctrl{1} & \qw & \qw  \\
    & \qw & \targ & \qw & \ctrlo{-1} & \qw & \targ \qw & \qw & \qw \\
    & \targ & \qw & \qw & \ctrlo{-1}& \qw & \qw  & \targ & \qw
    }}
    
    \caption{Circuit implementation of the QE evolution $U_{\alpha_2\beta_1\rightarrow\alpha_1\beta_2}$. The circuit for $U_{\alpha_1\beta_2\rightarrow\alpha_2\beta_1}$ is identical, but with the rotation angle flipped.}

\label{cir:qe2_implicit}
\end{figure}

We can then rewrite the circuits in Figs.~\ref{cir:qe1_implicit} and \ref{cir:qe2_implicit} in terms of single-qubit and CNOT gates. It was shown in Ref.~\cite{yordanov2020circuits} that a wisely chosen implementation of the multi-controlled rotation requires eight CNOTs, one of which cancels out with another one in the outer circuit. This results in circuits whose CNOT count of 13 matches that of the optimized product implementation of Refs.~\cite{Nam_2019,Wang_2021}, but whose CNOT depth is decreased to 11 (instead of 13). Figure~\ref{cir:qe1_explicit} shows the result of using this strategy to decompose the rotation in the circuit of Fig.~\ref{cir:qe1_implicit}. A similar strategy can be applied to the circuit in Fig.~\ref{cir:qe2_implicit}, and more generally to any double QE evolution.

\begin{figure*}[htbp]

    \footnotesize
    \centerline{
    \Qcircuit @C=0.2em @R=.7em {
    & \ctrl{2} & \qw & \ctrl{1} & \qw & \gate{R_y\left(\text{-}\frac{\theta}{4}\right)} & \ctrl{2} & \gate{R_y\left(\frac{\theta}{4}\right)} & \ctrl{3} & \gate{R_y\left(\text{-}\frac{\theta}{4}\right)} & \ctrl{2} & \gate{R_y\left(\frac{\theta}{4}\right)} & \ctrl{1} & \gate{R_y\left(\text{-}\frac{\theta}{4}\right)} & \ctrl{2} & \gate{R_y\left(\frac{\theta}{4}\right)} & \ctrl{3} & \gate{R_y\left(\text{-}\frac{\theta}{4}\right)} & \ctrl{2} & \gate{R_y\left(\frac{\theta}{4}\right)} & \ctrl{1} &\gate{S}& \ctrl{2} & \qw  & \qw\\
    & \qw & \ctrl{2} & \targ & \qw & \gate{H} & \qw & \qw & \qw & \qw & \qw & \qw & \targ &  \gate{R_y\left(\text{-}\frac{\pi}{2}\right)}  & \qw & \gate{S^\dagger}  &  \qw & \qw & \qw & \qw &\targ & \gate{S^\dagger} & \qw & \ctrl{2} & \qw\\
    & \targ & \qw & \qw & \gate{X} & \gate{H} & \targ & \qw & \qw & \qw & \targ & \qw & \qw & \qw & \targ & \qw & \qw & \qw & \targ  & \gate{H} & \gate{X} & \qw & \targ & \qw  & \qw\\
    & \qw & \targ & \qw & \gate{X} & \gate{H} & \qw & \qw & \targ & \qw & \qw & \qw & \qw & \qw & \qw & \qw & \targ & \gate{H} & \gate{X}
     & \qw & \qw & \qw & \qw & \targ & \qw}}
     
    \caption{Explicit implementation of the qubit excitation evolution $U_{\alpha_1\beta_1\rightarrow\alpha_2\beta_2}$.}

\label{cir:qe1_explicit}
\end{figure*}

We recall that we are considering the special case of double QEs acting on four spin-orbitals which are equally divided between $\alpha$-type and $\beta$-type. If all orbitals are of the same type, there are not two, but three unique double QEs: $T_{\alpha_1\alpha_3\rightarrow\alpha_2\alpha_4}^{(QE)}$, $T_{\alpha_1\alpha_4\rightarrow\alpha_2\alpha_3}^{(QE)}$, and $T_{\alpha_1\alpha_2\rightarrow\alpha_3\alpha_4}^{(QE)}$ for $\alpha$-type, and similarly for $\beta$-type. Their structure and circuit implementation can easily be found using the same methods. Assuming they are ordered as $\alpha_4$, $\alpha_3$, $\alpha_2$, $\alpha_1$ (little endian), the unitary generated by $T_{\alpha_1\alpha_3\rightarrow\alpha_2\alpha_4}^{(QE)}$ can be implemented as in Fig.~\ref{cir:qe1_implicit}, and the one generated by $T_{\alpha_1\alpha_4\rightarrow\alpha_2\alpha_3}^{(QE)}$ as in Fig.~\ref{cir:qe2_implicit}. None of the circuits directly implements  $T_{\alpha_1\alpha_2\rightarrow\alpha_3\alpha_4}^{(QE)}$, but the corresponding circuit implementation is easily derived from the others (e.g. we can simply exchange the roles of qubits 2 and 3 in Fig.~\ref{cir:qe1_implicit}). It is then straightforward to again obtain explicit circuits with a CNOT count of 13 and CNOT depth of 11.

Single and double QEs are the constituents of the pool used in QEB-ADAPT-VQE \cite{Yordanov_2021}, the most circuit-efficient ADAPT-VQE protocol to date. We have not discussed single QEs. Evidently, they only exist for pairs of spin-orbitals of the same type, and each such pair admits exactly one unique QE. The corresponding evolutions can be implemented using circuits with 2 CNOTs \cite{Yordanov_2021}.

\subsubsection{Multiple Variational Parameters (MVP)-CEOs}
\label{sss:mvp_ceo}

In the previous section we saw that each QE exchanges exactly two Slater determinants. In this section, we will define operators which are capable of doing all valid exchanges simultaneously. By `valid exchanges' we mean those that preserve particle number and $S_z$. In the example of the previous subsection, this would mean exchanging $\ket{0101}\leftrightarrow\ket{1010}$ \textit{and} $\ket{1001}\leftrightarrow\ket{0110}$.

As before, we consider a set of four spin-orbitals which is equally divided between $\alpha$- and $\beta$-type. We do not enforce restrictions on the spin-orbitals; they may or not be occupied in the reference state. We label them $\alpha_1$, $\beta_1$, $\alpha_2$, $\beta_2$ and define the following family of parameterized operators, consisting of linear combinations of the two unique QEs acting on these spin-orbitals:
\begin{align}
\begin{split}
&T_{\alpha_1\beta_1\alpha_2\beta_2}^{(MVP-CEO)}(\theta_1,\theta_2)=
\theta_1 T_{\alpha_1\beta_1\rightarrow\alpha_2\beta_2}^{(QE)} + \theta_2 T_{\alpha_2\beta_1\rightarrow\alpha_1\beta_2}^{(QE)}\\
&= \theta_1 Q^\dagger_{\alpha_2}Q^\dagger_{\beta_2}Q_{\alpha_1}Q_{\beta_1} + \theta_2 Q^\dagger_{\alpha_1}Q^\dagger_{\beta_2}Q_{\alpha_2}Q_{\beta_1} - h.c. 
\end{split}
\label{eq:ceo_3vp}
\end{align}
We note once again that the choice of whether to use, e.g., $T_{\alpha_1\beta_1\rightarrow\alpha_2\beta_2}^{(QE)}$ or $ T_{\alpha_2\beta_2\rightarrow\alpha_1\beta_1}^{(QE)}$ is irrelevant. A different choice might at most lead to a minus sign, which is absorbed by the variational parameters. Up to this irrelevant degree of freedom, the operator is unique. Thus, the unordered set of spin-orbital indices is enough to identify it unambiguously.

We call the operators of the type of Eq.~\eqref{eq:ceo_3vp} \textit{coupled exchange operators} (CEOs), because they combine the exchanges corresponding to multiple QEs in one operator. We use MVP (\textit{multiple variational parameters}) to indicate that the different exchanges are independently parameterized.  

The operator
\begin{equation}
U_{\alpha_1\beta_1\alpha_2\beta_2}^{(MVP-CEO)}(\theta_1,\theta_2)=e^{T_{\alpha_1\beta_1\alpha_2\beta_2}^{(MVP-CEO)}(\theta_1,\theta_2)}
\label{eq:mvp-ce_expon}
\end{equation}
will act as 
\begin{align}
\begin{split}
\ket{0101}&\rightarrow \cos{\theta_1}\ket{0101}+\sin{\theta_1}\ket{1010},\\
\ket{1010}&\rightarrow \cos{\theta_1}\ket{1010}-\sin{\theta_1}\ket{0101},\\
\ket{1001}&\rightarrow \cos{\theta_2}\ket{1001}+\sin{\theta_2}\ket{0110},\\
\ket{0110}&\rightarrow \cos{\theta_2}\ket{0110}-\sin{\theta_2}\ket{1001}\,,
\end{split}
\label{eq:ce_expon_exchanges}
\end{align}
where we again assume the qubits to be labeled $\alpha_2$, $\alpha_1$, $\beta_2$, $\beta_1$ from the most to the least significant. All other Slater determinants are left unchanged. Unlike QEs, these operators act non-trivially on \textit{all} Slater determinants where the underlying rotation preserves $S_z$ and particle number. In what concerns single excitations, we see that CEOs and QEs are identical---in this case, there is only one valid exchange.

The question that remains is how to create a quantum circuit which implements the coupled exchange evolution in Eq.~\eqref{eq:mvp-ce_expon}. Since QEs acting on the same set of spin-orbitals commute, we can implement this unitary by concatenating two circuits with a similar structure to the one in Fig.~\ref{cir:qe1_explicit}, resulting in a circuit with a total of 26 CNOTs. However, this is not optimal. Plugging the expressions in Eqs.~\eqref{eq:qe1} and \eqref{eq:qe2} into the definition of the CEOs, we see that they consist of a linear combination of eight Pauli strings:
\begin{align}
\begin{split}
&T_{\alpha_1\beta_1\alpha_2\beta_2}^{(MVP-CEO)}(\theta_1, \theta_2)
=\\
&\frac{i}{8}[
+(\theta_1 + \theta_2)XXXY - (\theta_1 + \theta_2)XXYX\\ 
& +(\theta_1 - \theta_2)XYXX + (\theta_1 - \theta_2) XYYY \\
&-(\theta_1-\theta_2)YXXX- (\theta_1 - \theta_2)YXYY\\
&+(\theta_1 + \theta_2)YYXY - (\theta_1 + \theta_2)YYYX].
\label{eq:ceo_paulis}
\end{split}
\end{align}

This happens because all QEs acting on the same set of spin-orbitals consist of uniformly weighted linear combinations of the same Pauli strings, with the difference residing in the signs of the coefficients.

The circuit implementation specific to QEs proposed in Ref.~\cite{yordanov2020circuits} does not apply to our CEOs, because it relies on the fact that all Pauli strings have the same weight. However, we can implement the unitary generated by the operator in Eq.~\eqref{eq:ceo_paulis} using the optimized circuits for exponentials of commuting Pauli strings proposed in Refs.~\cite{Nam_2019,Wang_2021}. In Fig.~\ref{cir:ceo_ind_params} we provide a 13-CNOT circuit which implements the exponential of any linear combination of the eight Pauli strings we are concerned with, and which implements the unitary $e^{T_{\alpha_1\beta_1\alpha_2\beta_2}^{(MVP-CEO)}(\theta_1, \theta_2)}$ as a special case with only two independent parameters.

\begin{figure*}[htbp]

    \footnotesize
    \centerline{
    \Qcircuit @C=0.3em @R=.7em {
    &\gate{S^\dagger} & \ctrl{3} & \ctrl{2} & \ctrl{1} & \gate{H} & \gate{R_z\left(-\theta_0\right)} & \targ & \gate{R_z\left(-\theta_1\right)} & \targ & \gate{R_z\left(\theta_2\right)} & \targ & \gate{R_z\left(-\theta_3\right)} & \targ & \gate{R_z\left(\theta_4\right)} & \targ & \gate{R_z\left(\theta_5\right)} & \targ & \gate{R_z\left(\theta_6\right)} & \targ & \gate{R_z\left(-\theta_7\right)} & \gate{H} & \ctrl{1} & \ctrl{2} & \ctrl{3} & \qw & \qw  \\
    & \qw & \qw & \qw & \targ & \qw & \qw & \ctrl{-1} & \qw & \qw & \qw & \ctrl{-1} & \qw & \qw & \qw & \ctrl{-1} & \qw & \qw & \qw & \ctrl{-1} & \qw & \qw & \targ & \qw & \qw  & \qw  & \qw \\
    & \qw & \qw & \targ & \qw &\gate{S^\dagger} & \qw & \qw & \qw & \qw & \qw & \qw & \qw & \ctrl{-2} & \qw & \qw & \qw & \qw & \qw & \qw & \qw & \qw & \qw & \targ & \qw &\gate{S} & \qw \\
    & \qw & \targ & \qw & \qw & \qw & \qw & \qw & \qw & \ctrl{-3} & \qw & \qw & \qw & \qw & \qw & \qw & \qw &\ctrl{-3} \qw & \qw & \qw 
    & \qw & \qw & \qw & \qw & \targ & \qw  & \qw }}
    
    \caption{Circuit implementation of $e^{\frac{i}{8} (\theta_0XXXY + \theta_1XXYX + \theta_2YXYY + \theta_3YXXX + \theta_4YYXY + \theta_5YYYX + \theta_6XYYY + \theta_7XYXX)}$.}

\label{cir:ceo_ind_params}
\end{figure*}

So far, we have focused on the case where we have four spin-orbitals which are equally divided between $\alpha$-type and $\beta$-type (often referred to as ``opposite-spin excitations''). If we instead we have a set of four spin-orbitals of the same type (``same-spin excitations''), we have not two, but three distinct QEs acting on this set. In this case, we have a CEO with three variational parameters, which simultaneously exchanges three different pairs of Slater determinants. Nevertheless, it still corresponds to a linear combination of the same eight Pauli strings, so that the corresponding unitary can be implemented by the 13-CNOT circuit of Fig.~\ref{cir:ceo_ind_params}.

Since two-qubit QEs realize the one and only viable determinant exchange for the corresponding set of spin-orbitals, they trivially belong to the MVP-CEO set.

\subsubsection{One Variational Parameter (OVP)-CEOs}
\label{sss:ovp_ceo}

In the preceding section we proposed CNOT-efficient circuits for MVP-CEOs. We will now show that it is possible to further decrease the CNOT count for specific values of the variational parameters $\theta_1$, $\theta_2$. Specifically, if the two variational parameters have the same magnitude, we obtain the following two subfamilies of CEOs:
\begin{align}
\begin{split}
&T_{\alpha_1\beta_1\alpha_2\beta_2}^{(OVP-CEO,+)}(\theta)=
\theta(T_{\alpha_1\beta_1\rightarrow\alpha_2\beta_2}^{(QE)} + T_{\alpha_2\beta_1\rightarrow\alpha_1\beta_2}^{(QE)}) \\
&= \theta(Q^\dagger_{\alpha_2}Q^\dagger_{\beta_2}Q_{\alpha_1}Q_{\beta_1} + Q^\dagger_{\alpha_1}Q^\dagger_{\beta_2}Q_{\alpha_2}Q_{\beta_1}) - h.c.
\end{split}
\label{eq:ceo_sum}
\end{align}
\begin{align}
\begin{split}
&T_{\alpha_1\beta_1\alpha_2\beta_2}^{(OVP-CEO,-)}(\theta)= \theta(T_{\alpha_1\beta_1\rightarrow\alpha_2\beta_2}^{(QE)} - T_{\alpha_2\beta_1\rightarrow\alpha_1\beta_2}^{(QE)})\\ 
&= \theta(Q^\dagger_{\alpha_2}Q^\dagger_{\beta_2}Q_{\alpha_1}Q_{\beta_1} - Q^\dagger_{\alpha_1}Q^\dagger_{\beta_2}Q_{\alpha_2}Q_{\beta_1}) - h.c.
\end{split}
    \label{eq:ceo_diff}
\end{align}
We call these operators OVP-CEOs because they each have one variational parameter. We additionally use the superscript ($+$/$-$) to distinguish the two subfamilies. Note that the labeling is arbitrary---the signs are exchanged if we use $T_{\alpha_1\beta_2\rightarrow\alpha_2\beta_1}^{(QE)}$ instead of $T_{\alpha_2\beta_1\rightarrow\alpha_1\beta_2}^{(QE)}$. It can be readily seen that these operators consist of linear combinations of only four Pauli strings:
\begin{align}
\begin{split}
&T_{\alpha_1\beta_1\alpha_2\beta_2}^{(OVP-CEO,+)}(\theta)=\\
&\frac{i\theta}{4}\left(XXXY-XXYX+YYXY-YYYX\right)=\\
&\frac{i\theta}{4}XXXY\left(1-IIZZ-ZZII+ZZZZ\right),
\label{eq:ce1_paulis}
\end{split}
\end{align}
\begin{align}
\begin{split}
&T_{\alpha_1\beta_1\alpha_2\beta_2}^{(OVP-CEO,-)}(\theta)=\\
&
\frac{i\theta}{4}\left(XYXX+XYYY-YXXX-YXYY\right)=\\
&\frac{i\theta}{4}XXXY\left(IZIZ-IZZI-ZIIZ+ZIZI\right).
    \label{eq:ce2_paulis}
\end{split}
\end{align}
As before, we define the corresponding unitaries as
\begin{equation}
U_{\alpha_1\beta_1\alpha_2\beta_2}^{(OVP-CEO,\pm)}(\theta)=e^{T_{\alpha_1\beta_1\alpha_2\beta_2}^{(OVP-CEO,\pm)}(\theta)}.
\label{eq:ovp-ce_expon}
\end{equation}
The terms in each operator commute and they can be organized such that two adjacent terms differ on exactly two qubits. Then, it is straightforward to see that, using once again the methods of Refs.~\cite{Nam_2019,Wang_2021}, the corresponding unitaries can be implemented with a total CNOT count and depth of 9.

We can further improve this if we understand \textit{why} the CEOs correspond to a sum of a lower number of Pauli strings. This happens because, as compared to QEs, they are conditional on a simpler function of the parities of subsets of qubit states. We saw that $T_{\alpha_1\beta_1\rightarrow\alpha_2\beta_2}^{(QE)}$ will only \textit{not} quench a Slater determinant under three conditions: (i) $\alpha_1$ and $\beta_1$ must have the same occupation number, (ii) equivalently for $\alpha_2$ and $\beta_2$, and (iii) the occupation numbers in (i) must be opposite to those in (ii). These three conditions give rise to the three-control rotation in Fig.~\ref{cir:qe1_implicit}. In contrast, the CEO $T_{\alpha_1\beta_1\alpha_2\beta_2}^{(OVP-CEO,\pm)}$ will not quench those Slater determinants if (i) the total occupation number of the $\alpha$ orbitals is exactly one, and (ii) equivalently for the $\beta$ orbitals. This observation leads us to the circuit implementations in Figs.~\ref{cir:ceo1_implicit} and \ref{cir:ceo2}.

\begin{figure}[htbp]

    \centerline{
    \Qcircuit @C=1em @R=.7em {
    & \ctrl{1} & \ctrl{2} & \gate{R_y(-2\theta)} & \ctrl{2} & \ctrl{1} & \qw \\
    & \targ & \qw & \ctrl{-1}& \qw & \targ & \qw  \\
    & \ctrl{1} & \targ & \qw & \targ & \ctrl{1} & \qw\\
    & \targ & \qw & \ctrl{-2}& \qw & \targ  & \qw 
    }}
    
    \caption{Circuit implementation of $U_{\alpha_1\beta_1\alpha_2\beta_2}^{(OVP-CEO,+)}(\theta)$.}

\label{cir:ceo1_implicit}
\end{figure}

\begin{figure}[htbp]
\centerline{
\Qcircuit @C=1em @R=.7em {
& \targ & \qw & \ctrl{2}& \qw & \targ  & \qw \\
& \ctrl{-1} & \targ & \qw & \targ & \ctrl{-1} \qw & \qw\\
& \targ & \qw & \ctrl{1}& \qw & \targ & \qw  \\
& \ctrl{-1} & \ctrl{-2} & \gate{R_y(2\theta)} & \ctrl{-2} & \ctrl{-1} & \qw
}}

    \caption{Circuit implementation of $U_{\alpha_1\beta_1\alpha_2\beta_2}^{(OVP-CEO,-)}(\theta)$.}
\label{cir:ceo2}
\end{figure}

To decompose the multi-controlled rotations, we follow a similar procedure to Ref.~\cite{yordanov2020circuits}. We implement them as in Fig.~\ref{cir:ccy}, convert the CZ gates into CNOT gates where the control is the 0th qubit, and use the identity in Fig.~\ref{cir:identity} to remove a CNOT. The final circuit for $U_{\alpha_1\beta_1\alpha_2\beta_2}^{(OVP-CEO,+)}$ is shown in Fig.~\ref{cir:ceo1_explicit}. A similar decomposition can be obtained for $U_{\alpha_1\beta_1\alpha_2\beta_2}^{(OVP-CEO,-)}$. With this, we see that coupled exchange evolutions can be implemented with a CNOT count of 9 and a CNOT depth of 7. This should be contrasted with the circuits for double QE evolutions, for which these values are 13 and 11, respectively. As compared to the circuits for the 4-qubit generators in the qubit pool \cite{Tang_2021}, with a CNOT count and depth of 6, the circuits we propose barely increase the CNOT depth. Yet, unlike them, our circuits fully conserve $S_z$ and particle number.

\begin{figure*}[htbp]

    \centerline{
    \Qcircuit @C=1em @R=.7em  @!R{
    & \gate{R_y(2\theta)} & \qw &&& \qw & \gate{R_y\left(\frac{\theta}{2}\right)} & \ctrl{2} & \gate{R_y\left(-\frac{\theta}{2}\right)} & \ctrl{1} & \gate{R_y\left(\frac{\theta}{2}\right)} & \ctrl{2} & \gate{R_y\left(-\frac{\theta}{2}\right)} & \ctrl{1} & \qw\\
    &\ctrl{-1}& \qw & \push{\rule{.3em}{0em}=\rule{.3em}{0em}} && \qw & \qw & \qw & \qw & \ctrl{0} & \qw & \qw & \qw & \ctrl{0} & \qw \\
    & \ctrl{-1} & \qw &&& \qw & \qw & \ctrl{0} & \qw & \qw & \qw & \ctrl{0} & \qw & \qw & \qw
    }}
    
    \caption{Decomposition of a two-control rotation. Any controlled gate can be used, as long as it imposes on the target (qubit 0) a reflection with respect to the Y axis, so that the sense of the rotation is inverted if the control is in state $\ket{1}$. We further note that the roles of the control qubits can be interchanged; this particular choice was made so that one controlled gate will be eliminated via the identity in Fig.~\ref{cir:identity}.}

\label{cir:ccy}
\end{figure*}

\begin{figure}[htbp]
\centerline{
\Qcircuit @C=1em @R=.7em @!R {
& \ctrl{1} & \qw & \ctrl{1} & \qw & & &\gate{S}& \qw & \ctrl{1} & \qw & \qw\\
& \targ & \gate{H} & \targ & \qw & \raisebox{2em}{=} & & \gate{R_y\left(-\frac{\pi}{2}\right)} & \gate{S^\dagger} & \targ & \gate{S^\dagger} & \qw 
}} 

    \caption{Equivalence used to reduce the CNOT count.}
\label{cir:identity}
\end{figure}

\begin{figure*}[htbp]

    \centerline{
    \Qcircuit @C=0.6em @R=.7em {
    & \ctrl{1} & \ctrl{2} & \gate{R_y\left(-\frac{\theta}{2}\right)} & \ctrl{3} & \gate{R_y\left(\frac{\theta}{2}\right)} & \ctrl{1} & \gate{R_y\left(-\frac{\theta}{2}\right)}& \ctrl{3} & \gate{R_y\left(\frac{\theta}{2}\right)} & \ctrl{2} & \gate{S} & \ctrl{1} & \qw & \qw \\
    & \targ & \qw & \gate{H} & \qw & \qw & \targ &\qw  & \qw & \gate{R_y\left(-\frac{\pi}{2}\right)} & \qw & \gate{S^\dagger} & \targ & \gate{S^\dagger} & \qw\\
    & \ctrl{1} & \targ & \qw & \qw & \qw & \qw & \qw  & \qw & \qw &\targ & \qw & \ctrl{1} & \qw & \qw\\
    & \targ & \qw & \gate{H} & \targ & \qw & \qw & \qw & \targ & \gate{H} & \qw & \qw & \targ & \qw & \qw
    }}
    
    \caption{Explicit implementation of the circuit in Fig.~\ref{cir:ceo1_implicit} using only CNOTs and single-qubit gates.}

    \label{cir:ceo1_explicit}
\end{figure*}

As before, we focused our analysis on the case where the spin-orbitals are equally divided between $\alpha$- and $\beta$- type. If all orbitals are of the same type, we consider all six unique OVP-CEOs that can be formed from sums and differences of pairs of these excitations. It is straightforward to generalize the circuit implementations. We additionally recall that all single QEs trivially belong to the OVP-CEO set.

\subsubsection{CEO-ADAPT-VQE Algorithm}
\label{sss:CEO-ADAPT}

We now propose the CEO-ADAPT-VQE algorithm, which makes use of the CEO operators defined in the previous subsections. This algorithm follows the usual ADAPT-VQE workflow (see Sec.~\ref{sss:adapt}), with modifications in steps \ref{step:meas_gradients} and \ref{step:selection}. The need for these modifications stems from MVP-CEOs having multiple variational parameters, unlike the operators of any previously proposed pool. A consequence of this is that there is no unique gradient to use as the selection criterion.

We begin by specifying the adopted notation. We use $\{T_k^{(OVP-CEO)}\}$ to denote the pool formed from the set of OVP-CEOs~\footnote{Note that we have defined the pool operators to include the variational parameters. This is because our operator selection step might append to the ansatz an unitary with multiple variational parameters. As such, our protocol does not fit into the usual framework where an unparameterized generator is multiplied by a single variational parameter.}. For each $A\in\{T_k^{(OVP-CEO)}\}$, we define a set 
\begin{equation}
M^{(QE)}(A)=\{B\in\{T_k^{(QE)}\}: \text{Supp}(B)=\text{Supp}(A)\},
\label{eq:twin_set}
\end{equation}
where `Supp' denotes the support of an operator (the set of qubits on which it acts non-trivially). $M^{(QE)}(A)$ is the set of QEs acting on the exact same spin-orbitals (qubits) as $A$. We further define $M^{(QE)}_{\neq 0}(A)$ as the set obtained from $M^{(QE)}(A)$ by removing all elements associated with zero energy derivatives. We use $\#$ to denote the cardinality of a set, and define a function $\texttt{max\_g}$ which takes an operator pool and selects the element generating the unitary with the highest gradient magnitude (at the point where the parameter is zero).\\

\textbf{Modified step \ref{step:meas_gradients}}:  
\begin{itemize}
    \item Evaluate the gradients of the unitaries generated by elements of the operator pool $\{T_k^{(OVP-CEO)}\}$. If the norm of the vector formed by these gradients is under a pre-defined convergence threshold $\epsilon$, terminate.~\footnote{This is the original step \ref{step:meas_gradients}, with the specification that the pool is the set of OVP-CEOs.}
\end{itemize}

\textbf{Modified step \ref{step:selection}}:  
\begin{itemize}
    \item Define $T_n^{(OVP-CEO)}\equiv\texttt{max\_g}(\{T_k^{(OVP-CEO)}\})$.
    \item If $\#M^{(QE)}_{\neq 0}(T_n^{(OVP-CEO)})=1$, add $e^{T_n^{(OVP-CEO)}}$ to the ansatz. 
    \item Otherwise, add $e^{T_n^{(MVP-CEO)}}$ to the ansatz, where 
    \begin{equation}
        T_n^{(MVP-CEO)}=\sum_{T_i^{(QE)}\in M^{(QE)}_{\neq 0}(T_n^{(OVP-CEO)})}\theta_i T_i^{(QE)}.
    \end{equation}
    Here $\theta_i$ is an independent variational parameter associated with the generator $T_i^{(QE)}$. Note that the new unitary contains either two or three variational parameters.
\end{itemize}

Note that step \ref{step:meas_gradients} implies the same number of measurements for CEO-, QEB- and Qubit-ADAPT-VQE, because these operators are all formed from the same set of Pauli strings. The cost is $\mathcal{O}(N^5)$, where $N$ is the number of spin-orbitals \cite{anastasiou2023}. However, the total measurement cost will depend on the number of iterations. Modified step 3 requires knowing the gradients of some QE evolutions, but such gradients are always linear combinations of gradients of OVP-CEO evolutions. Thus, all data required in step 3 of CEO-ADAPT-VQE was already collected in step 2.

In formulating a variant of ADAPT-VQE with CEOs, the main question is which operator to choose in each iteration. ADAPT-VQE typically chooses operators based on the gradient selection criterion; however, since MVP-CEOs do not have a unique variational parameter, this criterion does not straightforwardly apply to them. The criterion could be generalized to the sum of the gradient magnitudes of each independently parameterized constituent; for an MVP-CEO, this would correspond to the sum of the gradient magnitudes of all the QEs it contains, while for an OVP-CEO this would be the gradient magnitude as usual.

However, note that the gradient magnitude associated with $T_n^{OVP-CEO}$ is the sum of the gradient magnitudes of its constituent QEs. To see this, consider two OVP-CEOs acting on the same set of spin-orbitals, $T^{(OVP-CEO,+)}$ and $T^{(OVP-CEO,-)}$. By the triangle inequality, the gradient magnitude associated with each is bounded above by the sum of the magnitudes of the gradients associated with the constituent QEs. If the latter gradients have the same sign, $T^{(OVP-CEO,+)}$ saturates the bound; if the signs are opposite, $T^{(OVP-CEO,-)}$ does. Since $T_n^{OVP-CEO}$ is by definition the OVP-CEO associated with the highest gradient, its type (sum/difference) is the one which saturates the bound. Thus, the gradient associated with the selected OVP-CEO is the same as what we would obtain applying the generalized gradient criterion to the MVP-CEO formed from the same QEs. In summary, the three operators

\begin{enumerate}
    \item OVP-CEO with the highest gradient magnitude among all OVP-CEOs (labeled $T_n^{(OVP-CEO)}$ in step \ref{step:meas_gradients})
    \label{item:max_g_ovp}
    \item MVP-CEO whose QEs have the highest total sum of gradient magnitudes (`generalized gradient') among all MVP-CEOs
    \label{item:max_g_mvp_1}
    \item MVP-CEO acting on the same spin-orbitals as \ref{item:max_g_ovp}, minus the QEs with zero gradients (labeled $T_n^{(MVP-CEO)}$ in step \ref{step:meas_gradients}) \label{item:max_g_mvp_2}
\end{enumerate}

usually have equivalent gradients, and operators \ref{item:max_g_mvp_1} and \ref{item:max_g_mvp_2} are usually the same (this is not always the case due to the existence of MVP-CEOs with 3 QEs; however, the difference is, in general, not significant enough to impact the results). In light of this, we choose to consider \ref{item:max_g_mvp_2} as the leading candidate among all MVP-CEOs. 

The next step is to decide between $T_n^{(OVP-CEO)}$ and $T_n^{(MVP-CEO)}$, since we have a tie in (generalized) gradients. OVP-CEOs have the benefit of being implemented with fewer CNOTs, while MVP-CEOs offer more variational freedom; the question is how to balance these factors. We choose to allow a QE to have an individual variational parameter if and only if it has a nonzero gradient. This is based on the expectation that, as a general rule, operators with zero gradients have no impact on the energy regardless of the value of the variational parameter, which we confirm in numerical simulations. Thus, if we consider an OVP-CEO of the form $\theta(T_1^{(QE)}\pm T_2^{(QE)})$ where $T_2^{(QE)}$ has zero gradient in the current variational state, the corresponding evolution $e^{\theta(T_1^{(QE)}\pm T_2^{(QE)})}$ is equivalent to $e^{\theta T_1^{(QE)}}$ when acting on this particular state. Despite being equivalent in practice, the former unitary has a more efficient implementation (Fig.~\ref{cir:ceo1_explicit}) than the latter (Fig.~\ref{cir:qe1_explicit}), requiring 9 CNOTs instead of 13.

Let us take as a simple example the case of the $H_2$ molecule, represented by four qubits in a minimal basis set. We use little endian ordering and assume the spin-orbitals to be ordered 1-4, where the first two are $\beta$-type and the others are $\alpha$-type. At the beginning of the first iteration, we represent the state by $\ket{0101}$ (the Hartree-Fock solution). For this system, we have only two distinct double QEs, $T_{1,3\rightarrow 2,4}^{QE}$ and $T_{1,4\rightarrow 2,3}^{QE}$. The first one acts as
\begin{align}
\begin{split}
\ket{0101}&\rightarrow +\ket{1010},\\
\ket{1010}&\rightarrow -\ket{0101},
\end{split}
\label{eq:hf_ex_1}
\end{align}
while the second acts as 
\begin{align}
\begin{split}
\ket{1001}&\rightarrow +\ket{0110},\\
\ket{0110}&\rightarrow -\ket{1001}.
\end{split}
\label{eq:hf_ex_2}
\end{align}
As before, all other determinants are quenched. It is then clear that the unitary $e^{\theta T_{1,4\rightarrow 2,3}^{QE}}$ leaves the Hartree-Fock reference state unchanged for any value of $\theta$ (i.e., $e^{\theta T_{1,4\rightarrow 2,3}^{QE}}\ket{0101}=\ket{0101}$). While $e^{\theta(T_{1,3\rightarrow 2,4}^{QE}+T_{1,4\rightarrow 2,3}^{QE})}$ and $e^{\theta T_{1,3\rightarrow 2,4}^{QE}}$ are distinct unitaries, they are equivalent when acting on any state that is orthogonal to $\ket{1001}$ and $\ket{0110}$. Therefore, we can choose the one that has the most efficient circuit implementation. This choice will not impact the variational state.

This motivates us to choose an OVP-CEO as opposed to an MVP-CEO when only one of the constituent excitations has nonzero gradient; however, the question of which CEO is the best choice when multiple constituent QEs have nonzero gradients remains open. It could be beneficial to opt for $T_n^{(OVP-CEO)}$ instead of $T_n^{(MVP-CEO)}$ even when the gradients of the underlying excitations are both nonzero. $T_n^{(OVP-CEO)}$ contains the pair of same-support QEs for which the sum of gradient magnitudes is the highest. If the signs of the gradients are the same, the operator $T_n^{(OVP-CEO)}$ is of the form $T_n^{(OVP-CEO,+)}=\theta(T_1^{QE}+T_2^{QE})$; otherwise, it is of the form $T_n^{(OVP-CEO,-)}=\theta(T_1^{QE}-T_2^{QE})$. Note that the sign of the gradient dictates the sign of the local minimizer found by optimizing the corresponding parameter (they are opposite). Since both the sum and the difference are included in the OVP-CEO pool, the gradient selection naturally favors coupling pairs of same-support excitations in such a way that the sign of the parameter can be matched with the sign of the minimizer for both operators. Of course, the magnitude is constrained to be the same for both. We could choose $T_n^{(OVP-CEO)}$ instead of $T_n^{(MVP-CEO)}$ hoping that the fact that the former has a more efficient circuit implementation compensates for the fact that the QEs are restricted to have the same parameter magnitude. 

In Sec.~I of the Supplementary Information, we consider a variant of CEO-ADAPT-VQE where the choice between $T_n^{(OVP-CEO)}$ and $T_n^{(MVP-CEO)}$ is based on the energy change the corresponding unitaries are capable of producing, scaled by the CNOT count of the corresponding circuits. Surprisingly, despite making a choice that optimizes the energy change per unit CNOT in any given iteration, this energy-based decision criterion results in a higher CNOT count than the gradient-based one for the same error. This justifies our decision to always opt for $T_n^{(MVP-CEO)}$ as long as it contains more than one QE with nonzero gradient. We refer to Sec.~I of the Supplementary Information for a discussion concerning this seemingly unexpected result. For the sake of completeness, this section additionally considers variants of ADAPT-VQE which use either OVP- or MVP-CEOs exclusively. 

\subsection{Classical numerical simulations}
\label{ss:classical_sims}

In this section, we compare the proposed CEO-ADAPT-VQE algorithm with the previous most hardware-efficient variants of ADAPT-VQE: QEB-ADAPT-VQE \cite{Yordanov_2021} and Qubit-ADAPT-VQE \cite{Tang_2021} (see Sec.~\ref{sss:adapt}). We first consider three systems in the STO-3G basis set: LiH (12 qubits) and BeH$_2$ (14 qubits) as examples of realistic molecules, and linear H$_6$ (12 qubits) as a proxy for strongly correlated systems whose classical simulation is not viable. We additionally consider triangular H$_6$, which is expected to be more difficult to simulate classically due to spin frustration. We use a higher convergence threshold $\epsilon$ for the largest system due to the corresponding pools being significantly larger.

\subsubsection{Convergence Plots: CEO- vs QEB- vs Qubit-ADAPT-VQE}
\label{sss:convergence_plots}

\begin{figure*}[htbp]

    \includegraphics{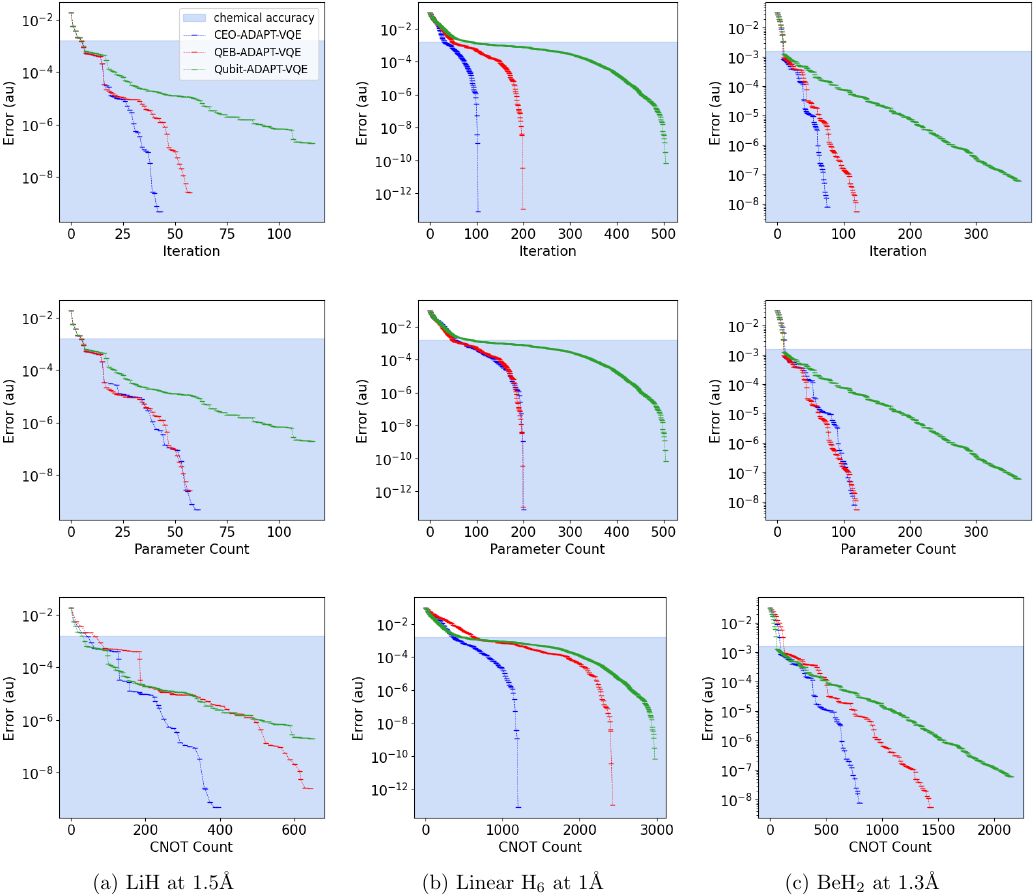}
    
     \caption{Convergence of the CEO-, QEB-, and Qubit-ADAPT-VQE algorithms for \textbf{a} LiH, \textbf{b} linear H$_6$, \textbf{c} BeH$_2$, at bond distances close to equilibrium. The error is plotted against the iteration number (top), parameter count (middle) and CNOT count (bottom). The region shaded blue is the region of chemical accuracy (error below 1kcal/mol). All algorithms terminate when the energy change falls below $10^{-10}$ Hartree.}
     
    \label{fig:main_equilibrium}
\end{figure*}

Figure~\ref{fig:main_equilibrium} shows the evolution of the three algorithms for three different molecules at bond distances close to equilibrium. In the upper panels, we can see that CEO-ADAPT-VQE is the fastest in decreasing the error with respect to the iteration number, which is equivalent to the number of optimizations and the number of gradient measurement rounds. 

In the middle panels, we see that for any given accuracy, CEO-ADAPT-VQE requires roughly the same number of parameters as QEB-ADAPT-VQE and significantly fewer than Qubit-ADAPT-VQE. Together with the lower number of iterations, these results show that among the three protocols, ours is the most frugal in terms of measurement costs. We recall that in each iteration, CEO-ADAPT-VQE adds up to three new variational parameters which, on average, will be associated with lower gradient magnitudes than either QEB- or Qubit-ADAPT-VQE. As such, it is a surprising result that it does not require a higher number of variational parameters to achieve similar accuracy. We expect this to be related to nonlocal effects, as it suggests that operators with low gradient magnitudes added earlier may affect the energy almost as much as operators with high gradient magnitudes added later on. It is plausible that an operator with a low gradient magnitude brings relevant Slater determinants into the superposition state, and/or interacts favorably with posteriorly added operators. As for the excess of variational parameters in Qubit-ADAPT-VQE as compared to CEO-/QEB-ADAPT-VQE, we attribute it to symmetry breaking. 

\begin{figure*}[htbp]

    \includegraphics{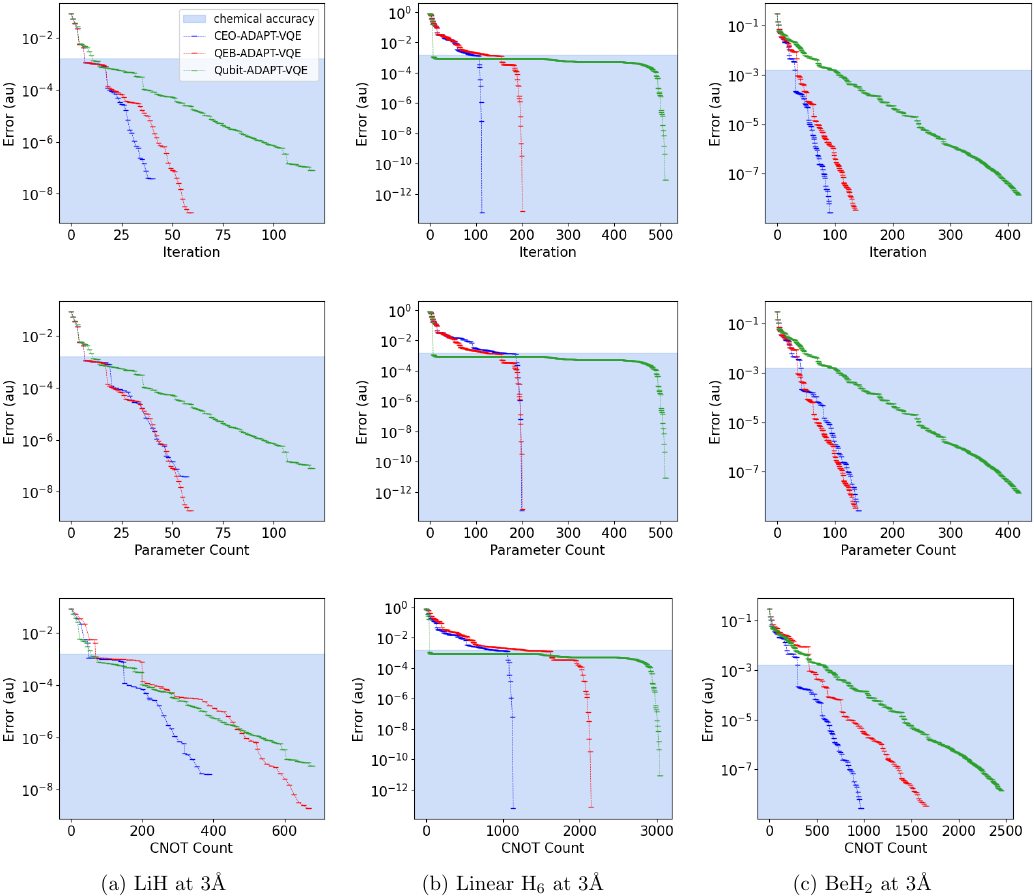}
    
     \caption{Convergence of the CEO-, QEB-, and Qubit-ADAPT-VQE algorithms for \textbf{a} LiH, \textbf{b} linear H$_6$, \textbf{c} BeH$_2$, at stretched bond distances. The error is plotted against the iteration number (top), parameter count (middle) and CNOT count (bottom). The termination criterion is the same as in Fig.~\ref{fig:main_equilibrium}.}
     
    \label{fig:main_stretched}
\end{figure*}

Finally, the bottom panels show the most important cost: the CNOT count. We observe that CEO-ADAPT-VQE achieves a significant reduction with respect to either of the other two algorithms. For linear H$_6$, Fig.~\ref{fig:main_equilibrium}(b), it requires at the moment of convergence roughly a third of the CNOTs of Qubit-ADAPT-VQE, and a half of QEB-ADAPT-VQE. Further, the error it has achieved upon reaching the termination criterion is at least as low as that of the other two algorithms. It is notable that such a reduction in the CNOT count can be achieved without any setbacks in regards to measurement costs or optimization difficulty (as gauged by the number of parameters).

In Fig.~\ref{fig:main_stretched} we consider the same molecules at stretched bond distances, where correlation effects should play a bigger role. Once more, we observe that CEO-ADAPT-VQE results in the most gate-efficient circuits among all three algorithms, without any drawbacks in terms of the number of parameters, number of optimizations, or measurement costs---on the contrary, these costs are reduced.

\begin{figure*}[htbp]

    \includegraphics{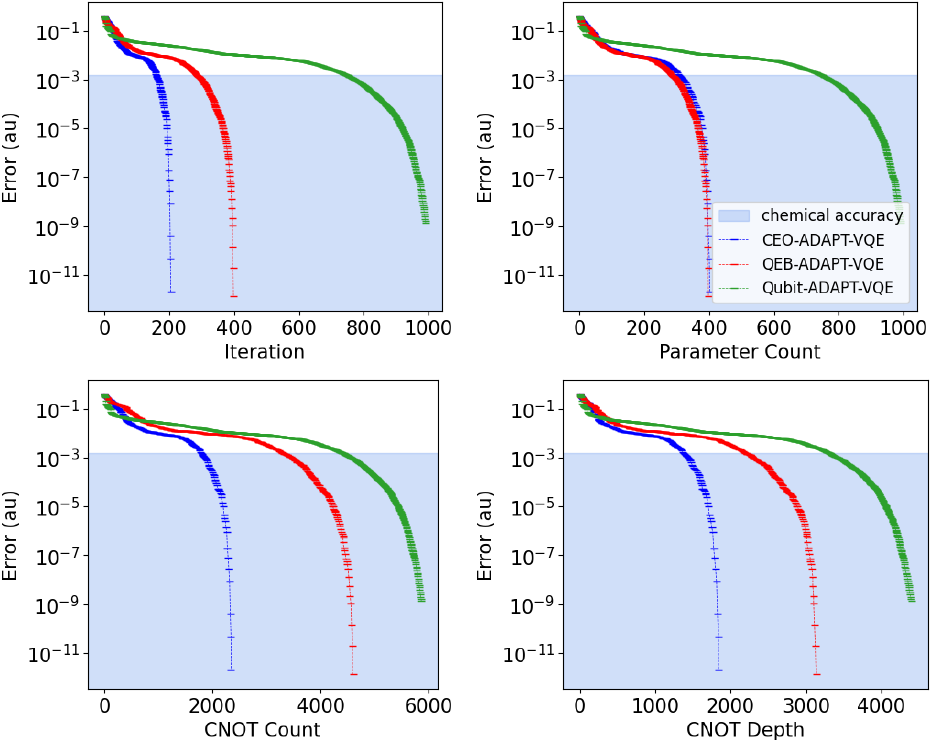}
    
     \caption{Convergence of the CEO-, QEB-, and Qubit-ADAPT-VQE algorithms for the triangular H$_6$ molecule at interatomic distance 2Å. This system is particularly challenging to simulate due to corresponding to a spin-frustrated lattice. The error is plotted against the iteration number (top left), parameter count (top right), CNOT count (bottom left), and CNOT depth (bottom right). The termination criterion is the same as in Fig.~\ref{fig:main_equilibrium}.}
    \medskip
     
    \label{fig:triangular_h6}
\end{figure*}

We note that for stretched linear H$_6$ (Fig.~\ref{fig:main_stretched}), Qubit-ADAPT-VQE is the best-performing variant up to the boundary of the chemical accuracy region. It may seem that, in this case, Qubit-ADAPT-VQE is the best option if the goal is reaching chemical accuracy faster; however, the state found by this algorithm at that point is actually a low-lying excited state with an energy roughly $10^{-3}$au higher than the ground state energy, which it has trouble steering away from for hundreds of iterations. Further, Qubit-ADAPT-VQE will generally converge faster in early iterations, where the state is likely less plagued by symmetry breaking (known to significantly affect convergence \cite{Bertels2022}). Because we are dealing with small, classically tractable systems, relatively few iterations suffice to reach chemical accuracy. However, a higher number of iterations will be required for larger molecules in general. Given that Qubit-ADAPT-VQE is typically outperformed by both CEO- and QEB-ADAPT-VQE after a few tens of iterations, we expect it not to be the leading algorithm for systems which are not amenable to classical simulation. In fact, the visible trend in which the difference in the CNOT counts becomes more significant for more strongly correlated molecules and larger iteration numbers indicates that CEO-ADAPT-VQE will be particularly advantageous for systems which are larger and/or harder to simulate classically.

Finally, Fig.~\ref{fig:triangular_h6} shows the convergence of the error obtained by the three adaptive algorithms against the iterations, CNOT count/depth and number of parameters for a triangular H$_6$ molecule, which exhibits spin frustration. Among all the systems considered here, this is the most challenging, which is confirmed by the fact that a larger number of iterations, CNOTs and parameters are required to reach the same convergence criterion. Yet, CEO-ADAPT-VQE clearly remains the most efficient option, requiring less than half of the CNOTs of the alternatives. As before, the parameter count is roughly the same for the CEO and QE pools, while being higher for the qubit pool. This confirms that our protocol can be successfully applied to strongly correlated systems, the expected use cases of quantum algorithms.

\subsubsection{Comparison with UCCSD-VQE through Bond Dissociation Curves}
\label{sss:bond_diss}

In this section, we compare CEO-ADAPT-VQE with (untrotterized) UCCSD-VQE for multiple values of the $\epsilon$ hyperparameter (the threshold on the gradient norm which defines convergence) on the energy, error and parameter count. 

\begin{figure*}[htbp]

    \includegraphics{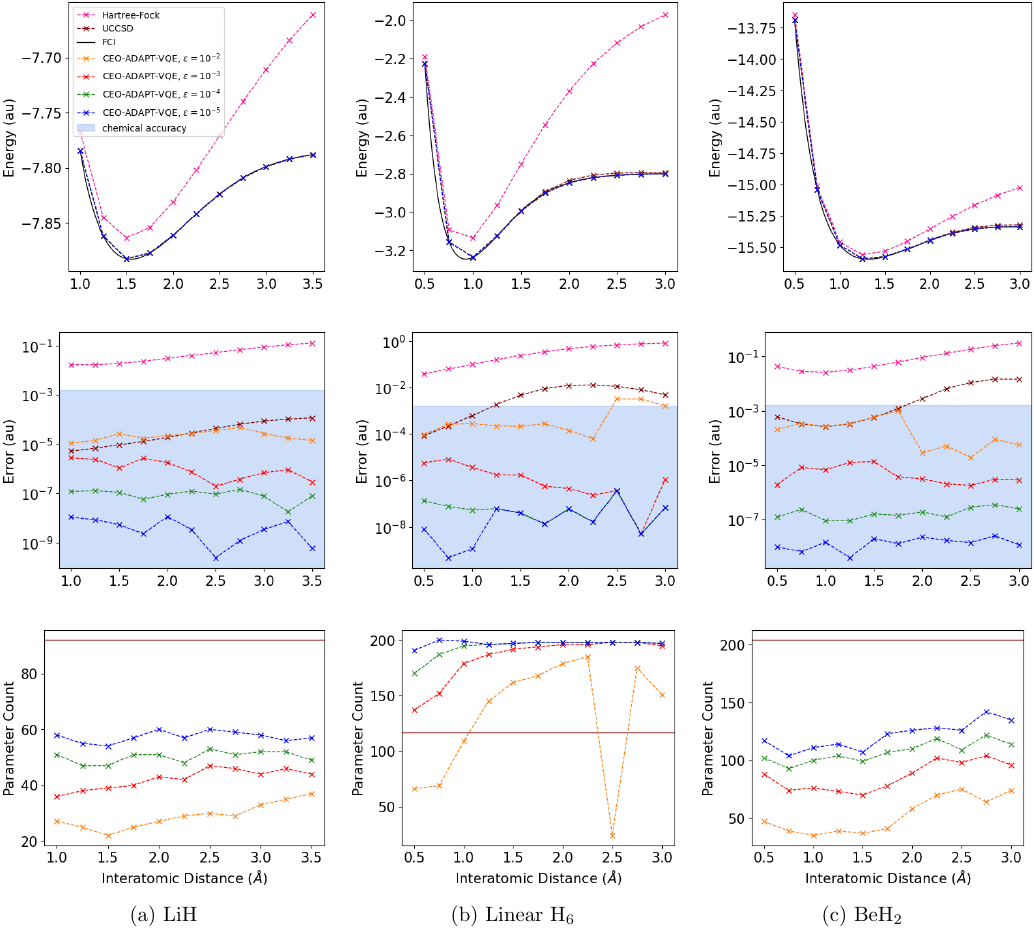}
    
    \caption{Energy, error and parameter count of CEO-ADAPT-VQE against the bond distance for \textbf{a} LiH, \textbf{b} linear H$_6$, \textbf{c} BeH$_2$. Four different values of the gradient convergence threshold $\epsilon$ are considered: $10^{-k}$ for $k\in\{2,3,4,5\}$. The Hartree-Fock, (untrotterized) UCCSD and FCI solutions are also included for comparison. The UCCSD ansatz includes all occupied-to-virtual single and double excitations. The corresponding parameter count is fixed at 92, 117, and 204 for the three molecules in the order that they appear.}
     
    \label{fig:bond_dissociation}
\end{figure*}

Figure~\ref{fig:bond_dissociation} shows these quantities as functions of the bond distance for all three molecules under study. In addition to UCCSD-VQE and CEO-ADAPT-VQE (with different values of $\epsilon$), we include the Hartree-Fock and FCI solutions for reference. We recall that the former is the starting point, while the latter is the target. 

In the upper and middle plots we observe that as expected, CEO-ADAPT-VQE always improves upon the Hartree-Fock solution by a significant margin. The middle plots show that as compared to UCCSD-VQE, it roughly matches or decreases the final error for all molecules and bond distances, and for all values of $\epsilon$ considered. Values of $\epsilon$ below $10^{-2}$ result in significant improvements with respect to UCCSD for all molecules and bond distances, with the error being lowered by a multiplicative factor of up to $10^6$.

In the bottom panels, we can see that our adaptive protocol produces more compact parameter vectors than UCCSD-VQE. For LiH (Fig.~\ref{fig:bond_dissociation}(a)) and BeH$_2$ (Fig.~\ref{fig:bond_dissociation}(c)), the parameter count is brought down by a factor of over 50\% on average, despite the increased accuracy. The trend is reversed for the case of H$_6$ (Fig.~\ref{fig:bond_dissociation}(b)), where CEO-ADAPT-VQE requires a larger parameter count for most bond distances---but in the middle panels, we can see that these are all cases where UCCSD-VQE fails to produce chemically accurate results, such that its performance is not satisfactory.

One more important observation is that lowering $\epsilon$ systematically results in a lower error. Thus, as compared to the other variants of ADAPT-VQE, CEO-ADAPT-VQE maintains the desirable feature that the final accuracy can be tuned by adjusting this hyperparameter. 

\subsubsection{ADAPT-VQE Evolution and State of the Art}
\label{sss:adapt_evolution}

The previous sections focused on benchmarking the performance of our new CEO variant of ADAPT-VQE. In order to showcase exclusively the improvements brought forth by the CEO pool, we did not consider any other proposals of improvements to ADAPT-VQE. However, several techniques to improve the hardware-efficiency and decrease the measurement costs of the algorithm are compatible with our protocol and can be readily incorporated into it. In this section, we consider an implementation of CEO-ADAPT-VQE which leverages such techniques, namely:

\begin{itemize}
    \item TETRIS: In this variant of ADAPT-VQE, multiple operators acting on disjoint qubit sets (i.e. with disjoint supports) are added to the ansatz in each iteration \cite{anastasiou2022}. Among all operators satisfying the disjoint support condition, priority is given to operators with higher gradient magnitudes. Once no operators with disjoint support and nonzero gradient remain, all new parameters are optimized and the iteration terminates.  This protocol promotes the creation of shallower circuits. Note than since we introduce a new dimension to the selection criterion, the magnitudes of the gradients of selected operators are expected to be lower on average. We could expect the average impact on the energy to be lower as a result. However, the original work showed that the CNOT count of the resulting circuits is not appreciably affected. TETRIS was applied to Qubit- and QEB-ADAPT-VQE by virtue of the fact that the corresponding operators act on at most four qubits (in contrast with fermionic operators, which act on a number of qubits which grows linearly with the size of the system \cite{Grimsley_2019}). This desirable characteristic is preserved by both types of CEOs, thus our proposed algorithm is similarly suited for the TETRIS protocol.
    \item Optimized gradient measurements (OGM): Given that the total number of distinct Pauli strings appearing in the QE/qubit pool and in the Hamiltonian both grow with $N^4$, we might naively expect the total number of strings in the commutators of the form of Eq.~\eqref{eq:adapt_commutator} to grow with $N^8$ for QEB/Qubit-ADAPT-VQE. However, it was shown that the pool strings can be grouped into sets of linear cardinality, leading to a grouping of all commutator strings into only $\mathcal{O}(N^5)$ commuting sets \cite{anastasiou2023}. Since CEO operators are linear combinations of QEs, they consist of the same set of Pauli strings, and thus this measurement strategy can be promptly generalized to our proposed pool. 
    \item Hessian recycling (HR): In the canonical implementation of ADAPT-VQE, each optimization collects from scratch information about the second-order derivatives of the cost function (either implicitly or explicitly). However, since the final state in a given iteration of the algorithm is taken as the initial state for the next, the local cost landscape at the end of a given optimization and at the beginning of the following one (restricted to shared parameters) is the same. In light of this, Ref.~\cite{ramôa2024} proposed to recycle between iterations the approximate inverse Hessian built iteratively by a quasi-Newton optimizer. An inverse Hessian of adequate dimension is constructed from the previous one by adding a new line and column which agree with the identity matrix. This inter-iteration exchange of curvature information was shown to improve the convergence rate of the optimization and reduce the total measurement costs of ADAPT-VQE for a variety of pools and molecules. Since this is a pool-agnostic method, it is straightforward to apply it to CEO-ADAPT-VQE. Note that the TETRIS protocol and MVP-CEOs both open the door to the possibility that the total number of cold-started variational parameters in a given optimization is greater than one. When several operators are added between two optimizations, no curvature information is known \textit{a priori} about any of them. Thus, the inter-iteration expansion of the inverse Hessian must in this case augment the matrix with multiple lines and columns.
\end{itemize}

Orbital optimization, as proposed in Ref.~\cite{Fitzpatrick_2024}, is also compatible with our proposed algorithm. However, while this technique may be advantageous when considering larger molecules and/or basis sets, its impact is negligible in the case of the minimal STO-3G basis set. Therefore, we do not consider it in this section. We refer to Sec.~III of the Supplementary Information for details about orbital optimization and its impact on CEO-ADAPT-VQE.

\begin{figure*}[htbp]

    \includegraphics{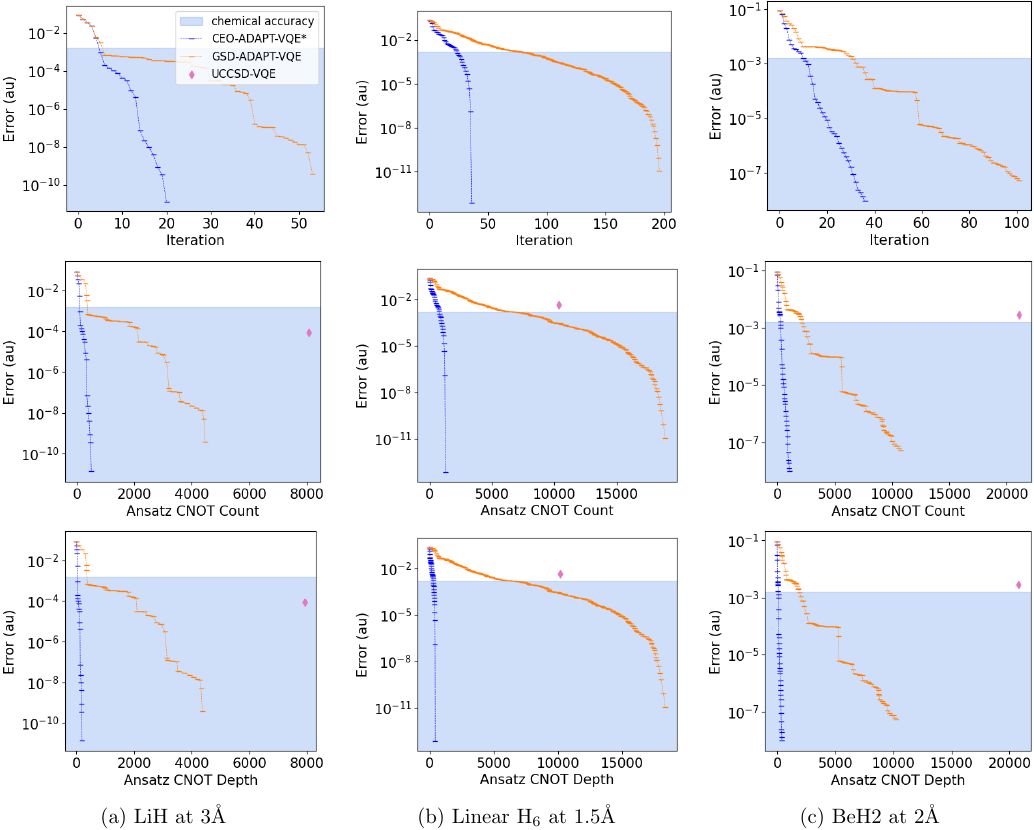}
    
    \caption{Convergence of the CEO- and GSD-ADAPT-VQE algorithms for \textbf{a} LiH, \textbf{b} linear H$_6$, \textbf{c} BeH$_2$, at various bond distances. The symbol $^*$ signals that the CEO-ADAPT-VQE algorithm represents the state of the art, being implemented \textit{in tandem} with other recent proposals: TETRIS \cite{anastasiou2022}, optimized gradient measurements \cite{anastasiou2023} and Hessian recycling \cite{ramôa2024}. In contrast, GSD-ADAPT-VQE follows the original protocol with a vanilla measurement strategy, representing the state of the art at the time ADAPT-VQE was first proposed \cite{Grimsley_2019}. The UCCSD ansatz is also included as a reference point. The corresponding circuit was implemented with one Trotter step and lexical ordering (with the excitations' source/target orbital indices as primary/secondary ordering criteria, respectively). The error is plotted against the iteration number, CNOT count, and CNOT depth. The region shaded blue is the region of chemical accuracy (error below 1kcal/mol). The convergence criterion is a gradient threshold of $10^{-6}$ and $10^{-5}$ on the 12-qubit and 14-qubit molecules, respectively.}
     
    \label{fig:combining_proposals_circuit}
\end{figure*}

\begin{figure*}[htbp]

    \includegraphics{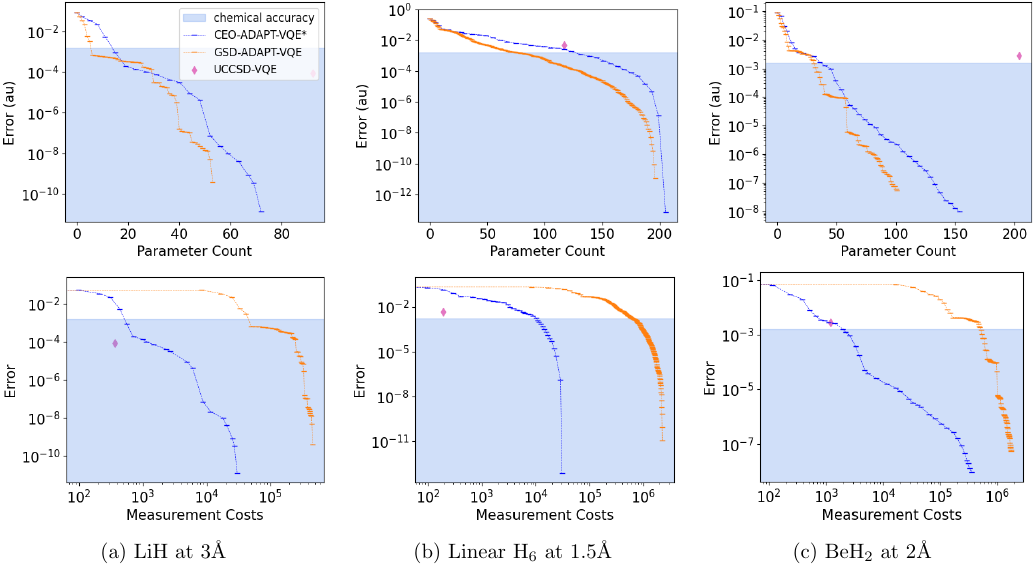}
    
    \caption{Convergence of the CEO- and GSD-ADAPT-VQE algorithms against the parameter count of the ansatz and measurement costs, which are given as multipliers for the cost of one energy measurement. The conditions are the same as in Fig.~\ref{fig:combining_proposals_circuit}.}
     
    \label{fig:combining_proposals_meas}
\end{figure*}

Figures \ref{fig:combining_proposals_circuit} and \ref{fig:combining_proposals_meas} compare the current state of the art of ADAPT-VQE against the original protocol. For an analysis of the individual impact of each of these strategies on CEO-ADAPT-VQE, we refer to Sec.~II of the Supplementary Information. In the following, we consider an algorithm implementing all strategies simultaneously: TETRIS-CEO-ADAPT-VQE enhanced with OGM and HR (which we label CEO-ADAPT-VQE* for simplicity). This hardware-efficient and cost-frugal variant of ADAPT-VQE, reflecting the improvements of recent years, is compared against GSD-ADAPT-VQE and UCCSD-VQE (implemented with lexical ordering and a single Trotter step). In the spirit of the original algorithm of Ref.~\cite{Grimsley_2019}, the latter algorithm uses a pool of generalized single and double fermionic excitations. In this case, we employ a vanilla approach where the gradient observables are not subject to grouping, such that the cost of step \ref{step:meas_gradients} scales with $\mathcal{O}(N^8)$. Further, we implement the circuits generated by GSD operators using ladders of CNOTs. In the case of doubles (which dominate costs), each circuit contains $16(n^{(df)}-1)$ entangling gates (where $n^{(df)}$ denotes the number of qubits the operator acts on nontrivially). Note that as of today, schemes exist for measuring the GSD gradients at $\mathcal{O}(N^6)$ cost \cite{Liu2021} and for implementing double fermionic excitation evolutions with $2n^{(df)}+5$ CNOTs \cite{yordanov2020circuits}. However, our implementation aims to reflect the knowledge at the time of the proposal of ADAPT-VQE.

It was shown in Ref.~\cite{anastasiou2023} that the cost of measuring the constituent strings of QEs at step \ref{step:meas_gradients} of ADAPT-VQE is upper bounded by $8N$ times the cost of one naive energy evaluation (for the same error). This is our estimate for the cost of this step in CEO-ADAPT-VQE with OGM. By similar arguments, the cost of this step when the measurements are done naively is $4N_s$, where $N_s$ is the number of distinct Pauli strings to be measured. This is our estimate for the cost of estimating the GSD gradients. 

In what concerns the energy, a naive protocol (i.e., independently measuring each Pauli string in the Hamiltonian) would lead to an $\mathcal{O}(N^4)$ evaluation cost. However, multiple strategies exist for grouping these strings into up to linear-sized commuting collections \cite{Gokhale2020,Verteletskyi_2020,Choi_2022,gokhale_2019_on3}. It is important to note that such groupings were shown in Ref.~\cite{McClean2016} to lead to additional covariance, such that the ratio of the number of commuting groups to the total number of strings is not straightforwardly related to the decrease in measurement costs. Therefore, it is not trivial to decide which is the best strategy, or how the costs relate to those of a naive Hamiltonian measurement for a given desired error. In both cases, the answer will be state- and Hamiltonian-dependent. A complete study is outside of the scope of our work. 
We consider the grouping based on $k$-commutativity proposed in Ref.~\cite{dalfavero2024}. This grouping has the advantage of allowing one to tune the depth of the measurement circuits: A lower value of $k$ leads to shallower circuits, while a higher value of $k$ leads to lower measurement costs. Given that the depth of the ansatz is expected to dominate over the (linear) depth of the measurement circuits \cite{Aaronson_2004,Gokhale2020meascircuits}, we take the maximum value $k=n$, where $n$ is the number of qubits (general commutativity). We include an analysis of the impact of $k$ on the number of collections and measurement costs for the systems under study in Sec.~IV of the Supplementary Information. 

Finally, based on the parameter-shift rules proposed in Refs.~\cite{Mitarai2018,Schuld_2019,Kottmann_2021}, we consider the cost of one gradient evaluation to be twice the cost of one energy evaluation.

To estimate the measurement cost reduction achieved by grouping the Hamiltonian relative to a no-grouping scenario, we consider the metric
\begin{equation}
    \hat{R}:=\left[\frac{\sum_{i=1}^N \sum_{j=1}^{m_i} \left| c_{ij} \right|}{\sum_{i=1}^N \sqrt{\sum_{j=1}^{m_i} \left| c_{ij} \right|^2}}\right]^2
    \label{eq:rhat}
\end{equation}
proposed in \cite{Crawford2021}. This metric gauges the savings achieved by grouping a Hamiltonian into $N$ collections, where collection $i$ has $m_i$ Pauli strings. $c_{ij}$ designates the coefficient of the $j$th string in the $i$th set. $\hat{R}$ approximates the expected value of the measurement cost ratio $M_u/M_g$, where $M_u$ and $M_g$ represent the shot count required to achieve a given accuracy with and without grouping, respectively, over the uniform spherical measure. This metric assumes that the distribution of measurements among collections is chosen so as to minimize the sampling error.

The uppermost panels in Fig.~\ref{fig:combining_proposals_circuit} reveal a significantly faster convergence of CEO-ADAPT-VQE* compared to GSD-ADAPT-VQE when we consider the number of iterations. While the iteration count is not a straightforward indicator of relevant costs, a lower number implies fewer measurement rounds and optimizations in total. The error against the CNOT count, plotted in the second row of panels, is particularly interesting to benchmark NISQ algorithms. We observe a remarkable reduction in the CNOT count required to reach a given error for all molecules. This reduction is the greatest for the most strongly correlated molecule (H$_6$). In this case, the improvement of CEO-ADAPT-VQE* with respect to GSD-ADAPT-VQE is even more significant than the improvement of the latter with respect to UCCSD-VQE. The third row of panels shows the error against the CNOT depth, which, given the short coherence times of NISQ devices, is often viewed as the go-to figure of merit to assess near-term viability. Owing to the TETRIS compactification, the reduction in CNOT depth achieved by CEO-ADAPT-VQE* is even greater than the reduction in CNOT count, with respect to the original algorithm. 

Figure \ref{fig:combining_proposals_meas} compares the parameter count (top) and measurement costs (bottom) of CEO-ADAPT-VQE* and GSD-ADAPT-VQE. We note that the former favors circuit-efficiency over parameter-efficiency; namely, MVP-CEOs and the TETRIS protocol are aimed at improving the energy reduction achieved by low-depth circuits without concern for the number of variational parameters. Yet, the number of additional parameters used in CEO-ADAPT-VQE* is very modest. Further, while a larger parameter vector is expected to cause the optimization to require more function and gradient evaluations due to being higher dimensional, the total measurement costs of CEO-ADAPT-VQE* are \textit{reduced} with respect to GSD-ADAPT-VQE. We attribute this to the use of the OGM and HR strategies, which significantly alleviate the overhead from the gradient measurement rounds and from the successive optimizations, respectively.

One more remarkable result we can observe in Fig.~\ref{fig:combining_proposals_meas} is that for the same error, the total measurement requirements of UCCSD-VQE are only significantly lower than those of CEO-ADAPT-VQE* for H$_6$, which we attribute to the phenomenon of gradient troughs known to plague this system \cite{Grimsley2023}. In this case, we observe a 29-fold increase in measurement costs (for the same accuracy), which nevertheless seems modest considering the widespread expectation of a prohibitive overhead and the fact that UCCSD-VQE is not a viable option for this system (given that it is not able to reach chemical accuracy). For the other systems, the measurement costs of CEO-ADAPT-VQE* are within an order of magnitude of those of UCCSD-VQE: They are increased only 4-fold for LiH, and are roughly matched for BeH$_2$. We note that the latter is the largest molecule, suggesting that CEO-ADAPT-VQE* may reduce measurement costs as we increase system size and approach classically intractable molecules, for which UCCSD has even more unnecessary parameters. This is a surprising result, considering that the measurement cost overhead incurred by the adaptive ansatz construction has been pointed out as a possible shortcoming of ADAPT-VQE, and a potential barrier to its practical viability. The reason for this expectation is two-fold: (i) the asymptotic cost of measuring the pool gradients required to adaptively build the ansatz is higher than the cost of measuring the energy, and (ii) ADAPT-VQE performs several optimizations per run, while static ans\"atze perform a single one. However, in practice, we observe that the total number of energy evaluations required by ADAPT-VQE throughout all optimizations until the UCCSD error is reached is actually \textit{lower} than the total required by the UCCSD-VQE optimization. This can be explained by the fact that growing the ansatz from scratch leads earlier optimizations to be lower dimensional, which in turn implies a faster convergence and cheaper gradient evaluations.  Additionally, it allows us to warm-start the parameters and inverse Hessian in each optimization. We can even view the sequence of ADAPT-VQE optimizations as a single adaptive optimization; such an optimization is particularly resource-efficient, because parameters are only included if or when they are actually relevant. Finally, we note that due to the lower parameter count and warm-starting of ADAPT-VQE optimizations, sampling noise is likely to be less detrimental for this algorithm than for UCCSD-VQE, which will further benefit the measurement costs of the former in a practical setting. This was demonstrated in Ref.~\cite{Ramoa2022}, where even the simplest implementation of the qubit-ADAPT-VQE algorithm (without TETRIS, OGM or HR) required only 5\% of the total shot count of UCCSD-VQE to reach chemical accuracy in the majority of the runs, when determining the ground state energy of molecular hydrogen (H$_2$) in the presence of finite sampling noise.

We remark that the improvements applied to CEO-ADAPT-VQE in this section could also be applied to QEB- and Qubit-ADAPT-VQE to obtain similar relative improvements in the CNOT depth and measurement costs.  The relative performance of the three algorithms would remain similar to what we observed in Sec.~\ref{sss:convergence_plots}, where none of the algorithms include the improvements. 

\subsubsection{Comparison with Fixed-Structure Ans\"atze}
\label{sss:adapt_vs_fixed}

In this section, we compare our adaptive algorithm (CEO-ADAPT-VQE*) against five leading non-adaptive ans\"atze for the electronic structure problem, namely:

\begin{itemize}
    \item Tiled Unitary Product State (tUPS) \cite{Burton_2024}
    \item Orbital-Optimized (oo)-tUPS \cite{Burton_2024}
    \item Perfect-Pairing (pp)-tUPS \cite{Burton_2024}
    \item Quantum Number Preserving (QNP) gate fabric \cite{Anselmetti_2021}
    \item Unitary Paired Coupled Cluster Generalized Singles and Doubles (k-UpCCGSD, where k indicates the number of layers) \cite{Lee_2018}
\end{itemize}

\begin{figure*}[htbp]

    \includegraphics{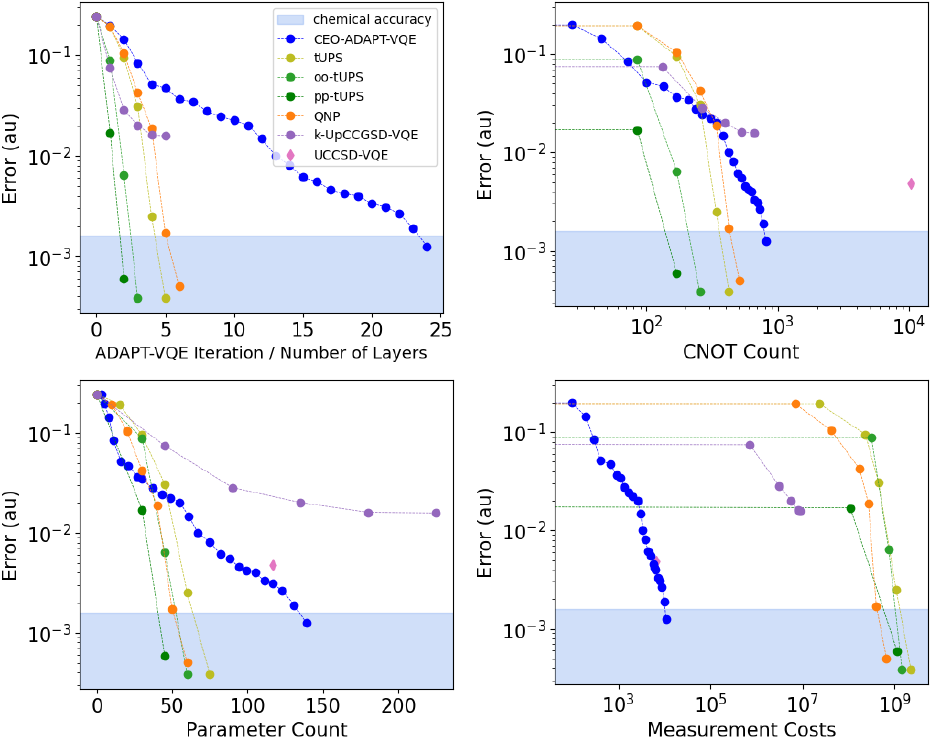}
    
    \caption{Convergence of CEO-ADAPT-VQE for linear H$_6$ at interatomic distance 1.5Å, as compared to a variety of state-of-the-art fixed-structure an\"satze: tUPS, oo-tUPs, pp-tUPS, QNP, and k-UpCCGSD-VQE. UCCSD-VQE (implemented with one Trotter step and lexical ordering) is also included for reference. The error is plotted against the number of layers (or iterations, in the case of ADAPT-VQE), CNOT count, parameter count, and measurement costs (given as multipliers for the cost of one energy measurement). The region shaded blue is the region of chemical accuracy (error below 1kcal/mol). All algorithms terminate when chemical accuracy is reached for the first time, except for k-UpCCGSD, the error of which stagnates before reaching this threshold.}
     
    \label{fig:ansatz_fight}
\end{figure*}

The optimization for the first four ans\"atze is done using a parallel-tempering basin-hopping approach \cite{Burton_2024}. In this method, local optimizations are followed by random perturbations of the parameters, such that multiple regions of the optimization landscape are searched for a good minimum. This optimization technique requires a significantly larger number of function evaluations than local optimization methods, and is aimed at local-trap-riddled cost landscapes. We note that the original QNP proposal in Ref.\cite{Anselmetti_2021} suggested two fixed-parameter initialization strategies followed by a local optimization; however, in our simulations, the QNP gate fabric did not result in satisfactory accuracy when we applied such strategies. In the case of the k-UpCCGSD ansatz, as suggested in the original work of Ref.~\cite{Lee_2018}, we employed 30 optimizations with parameters randomly initialized between -$\pi$ and $\pi$. For the CNOT count, we assumed this ansatz to be implemented using the fermionic swap networks proposed in Ref.~\cite{Ogorman_2019}.

The error from the basin-hopping optimizations is as provided in Ref.~\cite{Burton_2024}. The function evaluations were taken as the average over 10 optimizations, multiplied by the 8000 optimizations employed in the basin-hopping approach. This is due to the fact that Ref.~\cite{Burton_2024} did not make available the number of function/gradient evaluations, and the runtime of our Python implementation of the ansatz does not allow us to run thousands of optimizations in a reasonable time. As before, measurement costs include both energy and gradient evaluations, and all these evaluations are noiseless. We use the same optimizer (BFGS) and gradient convergence threshold ($10^{-8}$) in all cases.

Figure \ref{fig:ansatz_fight} compares the cost of the algorithms for linear H$_6$ at interatomic distance 1.5Å in terms of CNOTs, parameters, and measurement costs until chemical accuracy is reached. We observe that all algorithms succeed in reaching this threshold except for k-UpCCGSD, whose error stagnates after a few layers. While this could likely be improved by increasing the number of optimizations, the number of parameters and CNOTs suggest that this ansatz is not a competitive option, as they are higher than those of the other ans\"atze for a given error. The different variants of the tUPS ansatz and the QNP gate fabric succeed in reaching chemical accuracy, and do so with a lower CNOT count than CEO-ADAPT-VQE*. However, while the improvement in the CNOT count is of less than an order of magnitude, the measurement costs are increased by roughly 5 orders of magnitude. This is due to the difficulty of optimizing the rough parameter landscapes created by these tiled circuit structures. While ADAPT-VQE succeeds using parameters initialized at zero and local optimizations, these fixed-structure ans\"atze do not produce good results under such conditions. Hence, finding a good minimum requires performing many local optimizations, which increases the measurement costs to an impracticable extent. In addition to the number of energy evaluations being so high, no heuristics have been developed so far to determine how many optimizations are required to reach a good accuracy, and it is not known how the number scales with the size of the system. Therefore, despite the low CNOT counts, we do not expect these ans\"atze to constitute a viable option for realistic implementations.

\section{Discussion}
\label{s:discussion}

In this work, we proposed a new variant of ADAPT-VQE based on a pool of coupled exchange operators (CEOs) and showed that its performance is superior to any previously proposed variants of the algorithm. We further combined this protocol with other techniques aimed at decreasing the circuit depth and measurement costs of adaptively built ans\"atze. By uniting the CEO pool with optimized gradient measurement strategies \cite{anastasiou2023}, Hessian recycling \cite{ramôa2024} and the TETRIS protocol \cite{anastasiou2022}, we showed a total reduction of up to 88\%, 96\% and 99.6\% in the CNOT count, CNOT depth and measurement costs relative to the original ADAPT-VQE algorithm, for 12- to 14-qubit molecules. We additionally observed that in spite of the common belief that the adaptive ansatz construction incurs a significant measurement overhead, the total measurement costs of CEO-ADAPT-VQE* are actually comparable to those of UCCSD-VQE, a widely used static ansatz for the same problem. Further, we found numerical evidence of a decrease in the measurement costs of the former relative to the latter for larger molecules (Fig.~\ref{fig:combining_proposals_meas}), suggesting that our algorithm may actually offer a shot count reduction for classically intractable systems as compared to non-adaptive strategies. This conjecture was strengthened by a comparison with five leading fixed-structure ans\"atze, which were shown to either be unable to reach chemical accuracy or do so at the expense of a dramatic overhead in function evaluations as compared to our algorithm.

The new class of operators we introduced (CEOs) consists of linear combinations of qubit excitations (QEs) in which the QEs can be independently parameterized (MVP-CEOs) or share a single variational parameter (OVP-CEOs). We leveraged the structure of CEOs to implement the unitaries generated by them with similar or shallower circuits than those generated by QEs. We showed that such unitaries preserve particle number and spin symmetries, and that they can be implemented by circuits whose CNOT count is the same or lower as compared to any other circuits known to have these desirable symmetry preservation properties. Explicit CNOT-efficient circuit constructions were provided for all possible types of CEOs. We constructed circuits for MVP-CEO evolutions by optimizing a sequence of circuit implementations of exponentials of individual (commuting) Pauli strings, with a resulting CNOT count of 13---the same that is required to implement the evolution of a single one of its constituent QEs (which can be as many as three). In the case of OVP-CEO evolutions, we leveraged their structure as multi-controlled rotations to achieve an even lower CNOT count of 9.

The algorithm we propose, CEO-ADAPT-VQE*, is a variant of the quantum simulation algorithm ADAPT-VQE that makes use of the CEO operators to build a symmetry-adapted ansatz. We compared this algorithm with QEB- and Qubit-ADAPT-VQE, which to the best of our knowledge were the previous most hardware-efficient variants of ADAPT-VQE. Numerical simulations for various molecules show that CEO-ADAPT-VQE* significantly and systematically reduces the CNOT count of the ansatz as compared to these other variants. We found reductions by up to 65\%, with greater reductions for more strongly correlated molecules, which are also the most interesting systems---stronger correlations are harder to capture classically, thus making such systems possible candidates for quantum advantage experiments. We further observed that the reduction in the CNOT count increased with the number of algorithmic iterations, suggesting that larger molecules will favor CEO-ADAPT-VQE even more than our small, classically simulatable test cases.

The improved hardware-efficiency of the proposed adaptive VQE comes at no added cost. In fact, our algorithm requires \textit{fewer} iterations for the same accuracy than QEB- or Qubit-ADAPT-VQE. As such, it implies fewer optimizations and fewer gradient measurement rounds, effectively \textit{lowering} the measurement costs. What is more, the number of variational parameters it requires is either maintained or decreased with respect to other variants. 

Considering all the advantages it offers, we expect CEO-ADAPT-VQE* to be a leading candidate for molecular simulations on near-term quantum computers. In combination with other recent improvements to adaptive ans\"atze proposed in the literature, this algorithm is able to achieve a remarkable reduction in CNOT count/depth and measurement costs as compared to ADAPT-VQE at the time of its inception, showing great progress in the path towards quantum advantage with NISQ devices.

\section{Methods}
\label{s:methods}

Following the methods of Refs.~\cite{Grimsley_2019,Tang_2021,Yordanov_2021}, we compute the expectation values via matrix algebra, without including shot noise or hardware noise. This prevents confounding factors in the comparison of the performance of the algorithms  and allows for viable classical simulation runtimes. Since CNOT gates have the greatest physical error rates and error correction is out of bounds in the NISQ paradigm, we use the CNOT count of the circuits as the key figure to predict the impact of hardware noise in the algorithms \cite{Dalton2022}. Statistical noise is likely to significantly increase the measurement costs---hence, we expect our estimates to constitute a loose lower bound to the actual measurement costs in a realistic setting. The purpose of these estimates is not to provide an accurate estimation of these costs, but to get a sense of how they compare for the different algorithms.

We use the Openfermion \cite{openfermion} and PySCF \cite{pyscf} packages to create and manipulate fermionic operators. All errors are calculated with respect to the FCI energy. The classical optimizer employed on the VQE subroutine is BFGS as implemented in Scipy \cite{Scipy}, which we locally modified to recycle the Hessian in order to decrease the simulation time \cite{ramôa2024}.

\section*{Data Availability}

All data was generated with the code publicly available at \url{https://github.com/mafaldaramoa/ceo-adapt-vqe} and is available upon request.

\section*{Code Availability}

The code used for the numerical simulations has been made publicly available at \url{https://github.com/mafaldaramoa/ceo-adapt-vqe}.

\section*{Acknowledgments}

MR thanks Raffaele Santagati and Matthias Degroote for their support and encouragement at the start of this project, and Ernesto Galvão for helpful discussions. SEE and MR acknowledge support by Wellcome Leap as part of the Quantum for Bio Program. NJM and EB acknowledge support from the US Department of Energy (Grant No. 	DE-SC0024619). MR acknowledges support from FLAD (Luso-American Development Foundation) and from FCT (Fundação para a Ciência e a Tecnologia), under PhD research scholarship 2022.12333.BD.  This work is in part financed by National Funds through the Portuguese funding agency, FCT, within project LA/P/0063/2020.

\section*{Author Contributions}

M.R. conceptualized the CEO operators, created the corresponding circuits, developed the algorithm, and wrote the code for the simulations. P.G.A. conceived the MVP-CEO operators. All authors contributed to the development of the project, the discussion of the results and the writing of the manuscript.

\section*{Competing Interests}

The authors declare no competing interests.

\appendix 

\section{Variants of CEO-ADAPT-VQE}
\label{ap:ceo_variants}

CEO-ADAPT-VQE can be defined as a variant of ADAPT-VQE which uses a pool comprised of coupled exchange operators (CEOs). We defined two types of such operators: OVP- and MVP-CEOs, with one or up to three variational parameters respectively. One key decision in the algorithm is how to choose between these subsets in each iteration. In this section, we compare four different decision criteria.

As explained in the main text, in each iteration of our algorithm we use the gradients of OVP-CEOs to select two candidates: $T_n^{(OVP-CEO)}$, the OVP-CEO with the highest gradient, and $T_n^{(MVP-CEO)}$, the MVP-CEO formed from all QEs with nonzero gradients acting on the same spin-orbitals as $T_n^{(OVP-CEO)}$. One of these operators must then be chosen to generate the new ansatz unitary. 

We note that often, the gradient of $T_n^{(OVP-CEO)}$ is the same as the sum of the gradient magnitudes of the excitations in $T_n^{(MVP-CEO)}$. This is only not the case when the spin-orbitals are all of the same type, in which case the gradient of $T_n^{(MVP-CEO)}$ may be higher by virtue of including one more operator. Hence, we cannot in general use the gradient to decide between the two operators. One might expect that OVP-CEOs favor hardware-efficiency, since the corresponding evolutions are implemented with 9 CNOTs while MVP-CEOs are implemented with 13. However, the latter offer more variational freedom by implementing up to three independent qubit excitations each.

\begin{figure*}[htbp]

    \includegraphics{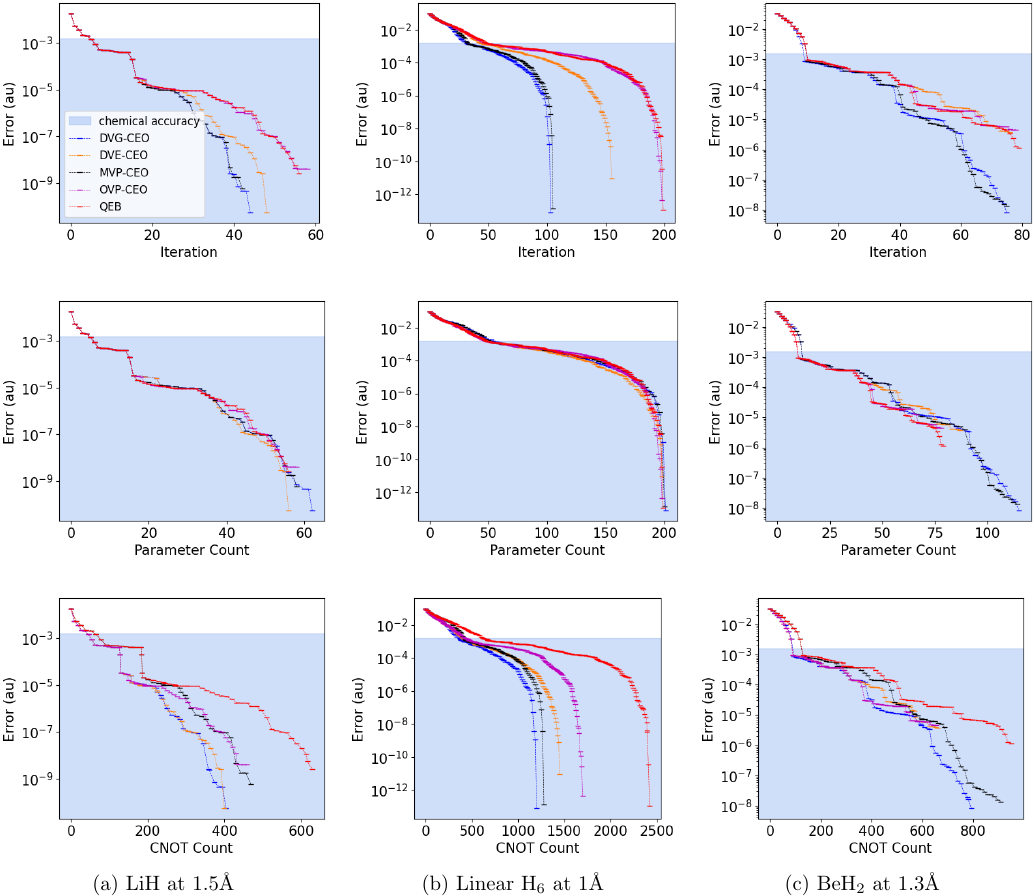}
    \caption{Comparison of CEO-ADAPT-VQE Variants at Equilibrium Bond Distances}
    \medskip
    \small
     The subfigures depict the convergence of different variants of the CEO-ADAPT-VQE algorithm for \textbf{a} LiH, \textbf{b} H$_6$, \textbf{c} BeH$_2$, at bond distances close to equilibrium. The QEB-ADAPT-VQE algorithm is also included for reference. The error is plotted against the iteration number (top), parameter count (middle) and CNOT count (bottom). The convergence criterion is a gradient threshold of $10^{-6}$ and $10^{-5}$ on the 12 qubits and 14 qubit molecules, respectively.
     
    \label{fig:ap_equilibrium}
\end{figure*}

We use the notation of the main text to define four variants of CEO-ADAPT-VQE which serve the purpose of assessing different criteria for this decision. Step 2 is left unchanged with respect to the one defined in the main text. In what concerns step 3, the definitions of $T_n^{(MVP-CEO)}$ and $T_n^{(OVP-CEO)}$ remain the same, but the criteria of choice between the two vary across variants as indicated below.

\begin{itemize}
    \item OVP-CEO-ADAPT-VQE: Add $e^{T_n^{(OVP-CEO)}}$ to the ansatz.
    \item MVP-CEO-ADAPT-VQE: Add $e^{T_n^{(MVP-CEO)}}$ to the ansatz.
    \item Decision via gradient (DVG)-CEO-ADAPT-VQE: Add $e^{T_n^{(OVP-CEO)}}$ to the ansatz if $\#M_{\neq 0}^{(QE)}(T_n^{(OVP-CEO)})=1$. Otherwise, add $e^{T_n^{(MVP-CEO)}}$.
    \item Decision via energy (DVE)-CEO-ADAPT-VQE: Add $e^{T_n^{(OVP-CEO)}}$ to the ansatz if $\#M_{\neq 0}^{(QE)}(T_n^{(OVP-CEO)})=1$. Otherwise, obtain $\Delta E_{OVP}$ and $\Delta E_{MVP}$, the energy changes produced by adding to the ansatz $e^{T_n^{(OVP-CEO)}}$ and $e^{T_n^{(MVP-CEO)}}$ (respectively) and performing a full optimization. Add the former unitary if $\frac{\Delta E_{MVP}}{13}>\frac{\Delta E_{OVP}}{9}$, and the latter otherwise~\footnote{Note that this energy change is expected to be negative.}.
\end{itemize}

In essence, we have two algorithms which use OVP-CEOs or MVP-CEOs exclusively, and two algorithms which combine the two operator types. DVG-CEO-ADAPT-VQE is the algorithm defined in the main text, where the decision is gradient-based: $T_n^{(MVP-CEO)}$ is chosen unless there is only one operator with nonzero gradient. In DVE-CEO-ADAPT-VQE, the decision is energy-based: We perform two independent optimizations of the ansatz with each of the two candidate unitaries, and effectively select the one which leads to the highest absolute value of the energy change per unit CNOT. The aim is to maximize the impact on the energy with respect to the added CNOT count in each iteration. Naturally, we could consider other hardware-related criteria, such as the CNOT depth. Note that this decision criterion roughly doubles the number of optimization required per ADAPT-VQE iteration.

Figure~\ref{fig:ap_equilibrium} compares these four algorithms. QEB-ADAPT-VQE is also plotted for reference. We verify that all variants of CEO-ADAPT-VQE improve upon QEB-ADAPT-VQE in terms of the number of CNOTs required for a given error, while being roughly matched in terms of the parameter count.

The decision via gradient performs the best among all variants of CEO-ADAPT-VQE, outperforming all others for all systems. This is remarkable, given the fact that the decision via energy is locally optimal (with respect to the CNOT count). In fact, in the first iteration where DVG- and DVE-CEO-ADAPT-VQE diverge, the latter is sure to have a lower error, because the optimization of the former is essentially a restricted version of the optimization of the latter. Yet, as the iterations proceed, the DVE variant lags behind. This can only be explained as a nonlocal effect. We hypothesize that the fact that DVG-CEO-ADAPT-VQE always includes all independently parameterized QEs with nonzero gradients leads to a higher variational flexibility which becomes more advantageous as the iterations proceed. Certain operators may have a low impact on the energy upon being added, but be beneficial later on --- perhaps they introduce important Slater determinants into the superposition state, or they interact favorably with operators added after them.

In general, the dynamic strategies (DVG and DVE), which decide between CEO types on a case-by-case basis, seem to outperform the predetermined ones. This is expected, since the latter make no effort to optimize the choice of operator type. However, there is a notable exception: MVP- beats DVE-CEO-ADAPT-VQE for H$_6$. Once again, this suggests that privileging extra variational freedom can be rewarding in the long term, and particularly so for highly correlated system---which strengthens the conjecture in the paragraph above.

\begin{figure*}[htbp]

    \includegraphics{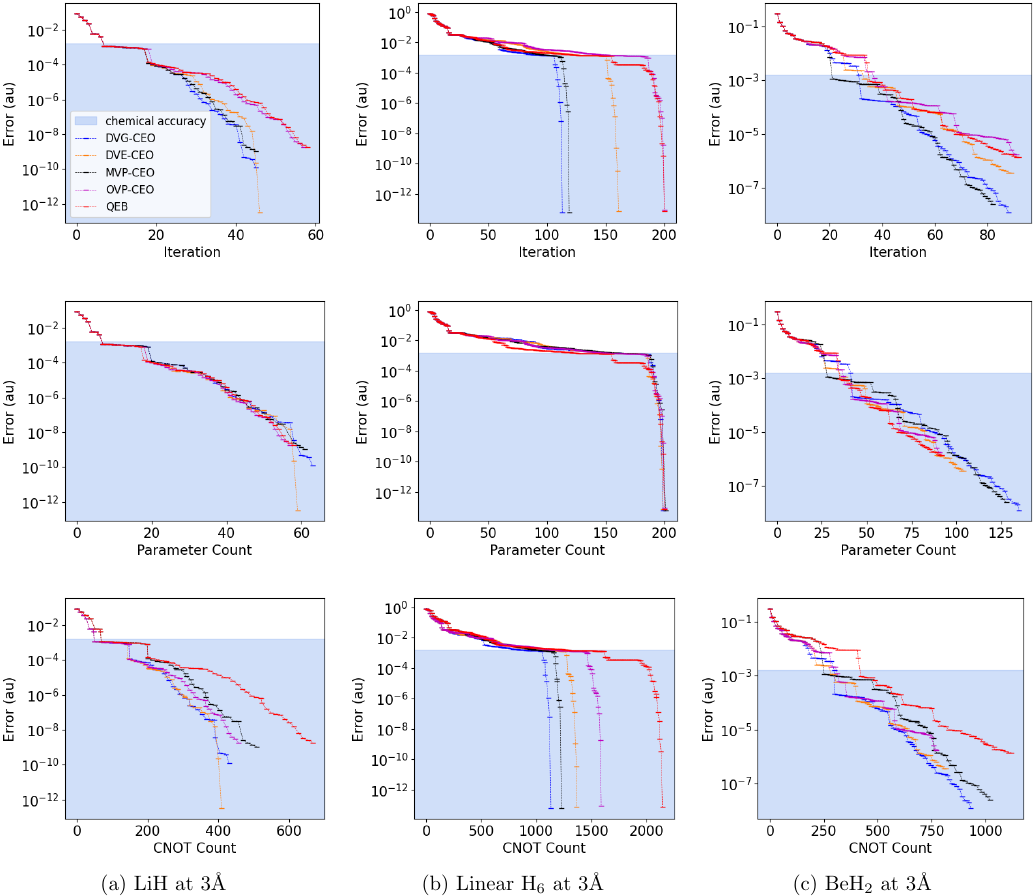}
    \caption{Comparison of CEO-ADAPT-VQE Variants at Stretched Bond Distances}
    \medskip
    \small
     Th subfigures depict the convergence of different variants of the CEO-ADAPT-VQE algorithm for \textbf{a} LiH, \textbf{b} H$_6$, \textbf{c} BeH$_2$, at stretched bond distances. The QEB-ADAPT-VQE algorithm is also included for reference. The error is plotted against the iteration number (top), parameter count (middle) and CNOT count (bottom). The convergence criterion is the same as in Fig.~\ref{fig:ap_equilibrium}.
     
    \label{fig:ap_stretched}
\end{figure*}

Figure~\ref{fig:ap_stretched} shows similar plots for stretched bond distances. We observe the same trends: QEB-ADAPT-VQE is outperformed by all variants of CEO-ADAPT-VQE, with the leading one being DVG. The parameter count is roughly equivalent for all five algorithms.

It has become evident that the decision via gradient is the optimal choice in terms of the CNOT count and number of iterations, despite requiring no extra optimizations as compared to the canonical ADAPT-VQE (unlike the decision via energy). Therefore, we take DVG-CEO-ADAPT-VQE as the standard ADAPT-VQE algorithm with CEOs and refer to it simply as CEO-ADAPT-VQE in the main text.

\section{Combining CEO-ADAPT-VQE with Transversal Proposals}
\label{ap:combining_proposals}

In this section, we investigate the individual impact of each of the strategies consolidated to create the CEO-ADAPT-VQE* algorithm: TETRIS \cite{anastasiou2022}, optimized gradient measurements (OGM) \cite{anastasiou2023} and Hessian recycling (HR) \cite{ramôa2024}. We refer to the main text for details about the methods.

\begin{figure*}[htbp]

    \includegraphics{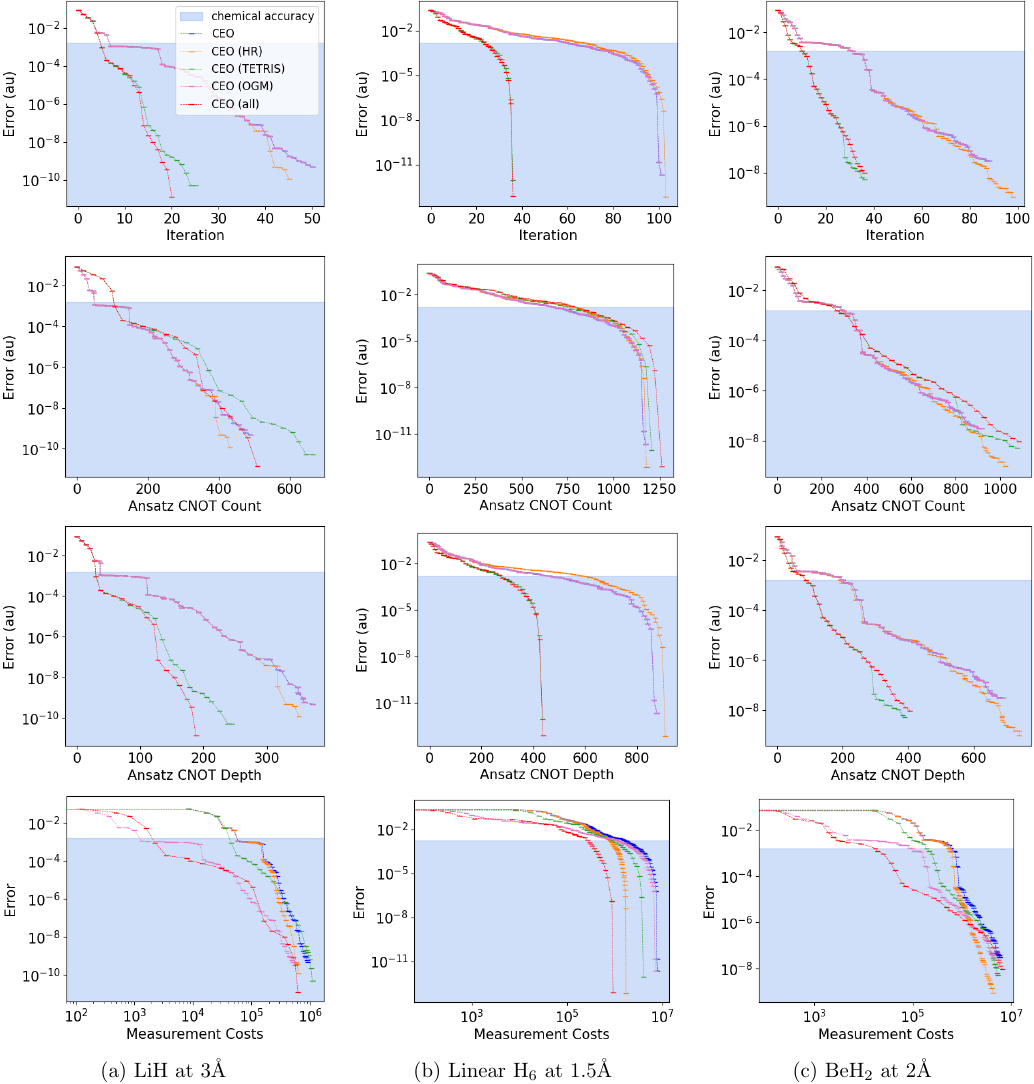}
    \caption{Combining CEO-ADAPT-VQE with Transversal Improvements}
    \medskip
    \small
    
     The subfigures depict the convergence of the CEO-ADAPT-VQE algorithm for \textbf{a} LiH, \textbf{b} H$_6$, \textbf{c} BeH$_2$, at various bond distances. The algorithm is implemented \textit{in tandem} with other recent proposals: Hessian recycling (HR) \cite{ramôa2024}, TETRIS \cite{anastasiou2022}, optimized gradient measurements (OGM)\cite{anastasiou2023}, and all three (all). The baseline case, which uses the CEO pool while following the original ADAPT-VQE protocol \cite{Grimsley_2019} with a vanilla measurement strategy, is also included for reference. The error is plotted against the iteration number, CNOT count, CNOT depth, and measurement costs. The region shaded blue is the region of chemical accuracy (error below 1kcal/mol). The convergence criterion is a gradient threshold of $10^{-6}$ and $10^{-5}$ on the 12 qubits and 14 qubit molecules, respectively. The curves for CEO and CEO (OGM) overlap on all plots except those pertaining to measurement costs.
    
    \label{fig:indep_combining_proposals}
\end{figure*}

Figure~\ref{fig:indep_combining_proposals} compares the convergence of CEO-ADAPT-VQE with no improvements, with all improvements, and with each improvement individually. The uppermost panels show that, as would be expected, only TETRIS relevantly affects the iteration count. The remaining strategies affect the total number of measurements per iteration, but leave the iteration count unchanged. The CNOT count (second row of panels) is similar for all curves, while the CNOT depth (third row) is predictably lowered by TETRIS while being roughly unchanged by the the other proposals. Finally, the last row of panels shows that OGM and HR contribute to decreasing the measurement costs of the algorithm. Interestingly, the relative impact of the two is system-dependent. While OGM is more impactful than HR for LiH and, in earlier iterations, for BeH$_2$, the reverse happens for H$_6$ and  BeH$_2$ in later iterations. We attribute this to the complexity of the optimizations. As an example, in the case of H$_6$, not only are the optimizations higher dimensional on average, but they also tend to require more cost function evaluations than optimizations for other molecules (even for matched parameter counts). As such, in this case, the cost of the energy measurements throughout the optimizations prevails over the cost of the gradient measurements, such that a strategy which tackles the cost of the optimization (HR) is more beneficial than one which tackles the cost of the gradient measurement round (OGM). We note that while its focus is decreasing circuit depth, TETRIS also offers a slight reduction in measurement costs by virtue of requiring a lower number of iterations (and thus fewer gradient measurement rounds and fewer optimizations in total).

Finally, we remark that the benefits of HR are expected to take more iterations to be harvested when we employ strategies which increase the number of new variational parameters per iteration, such as MVP-CEOs and TETRIS, due to the higher number of cold-started entries in the inverse Hessian. In the case of the 12 qubit molecules, TETRIS-CEO-ADAPT-VQE can add up up to three MVP-CEOs per iteration. Up to two of them may act on spatial orbitals of the same type, and thus have 3 variational parameters. In total, this may lead to up to 8 variational parameters, and the count will be higher for larger molecules. In early iterations, a large number of new parameters represents a significant number of second derivatives about which the recycled Hessian contains no information. Therefore, we can expect HR to become more impactful for larger and more strongly correlated systems, where the number of old parameters far outweighs the number of freshly added parameters in later iterations. 

\clearpage
\section{Orbital Optimization}
\label{ap:orbital_opt}

\begin{figure}[htbp]

    \includegraphics{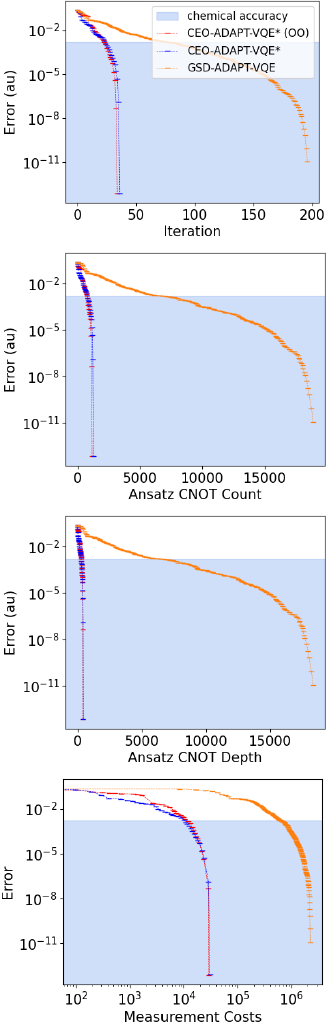}
    \caption{Combining CEO-ADAPT-VQE with Orbital Optimization}
    \medskip
    \small
    
     The subfigures depict the convergence of CEO-ADAPT-VQE*, as defined in the main text, with and without orbital optimization. GSD-ADAPT-VQE is also included for reference. We consider H$_6$ at an interatomic distance of 1.5\AA{}. CEO-ADAPT-VQE*, GSD-ADAPT-VQE and UCCSD-VQE are defined in the main text. The error is plotted against the iteration number, CNOT count, CNOT depth, and measurement costs. The measurement costs are given as multipliers for the cost of one energy measurement. The region shaded blue is the region of chemical accuracy (error below 1kcal/mol). The convergence criterion is a gradient threshold of $10^{-6}$ and $10^{-5}$ on the 12-qubit and 14-qubit molecules, respectively.
    
    \label{fig:H6_r=1.5_oo}
    
\end{figure}
 
Reference \cite{Fitzpatrick_2024} proposed ADAPT-VQE-SCF, an algorithm that merges orbital optimization techniques with ADAPT-VQE.

Given a variational state $\ket{\psi(\pmb{\theta})}$, we can obtain an orbital-optimized version $\ket{\tilde{\psi}(\pmb{\kappa},\pmb{\theta})}$ as

\begin{equation}
    \ket{\tilde{\psi}(\pmb{\kappa},\pmb{\theta})} = e^{\pmb{\kappa}}\ket{\psi(\pmb{\theta})}.
    \label{eq:oo_state}
\end{equation}

Here, $\pmb{\kappa}$ is an anti-hermitian operator defined by 

\begin{equation}
    \pmb{\kappa} = \sum_{p>q} \kappa_{pq} ( \hat{E}_{pq} - \hat{E}_{qp}),
    \label{eq:oo_gen}
\end{equation}

where the $\hat{E}_{pq}$ are spin-adapted one-body operators,

\begin{equation}
    \hat{E}_{pq} = \hat{a}^\dagger_{p\alpha}\hat{a}_{q\alpha} + \hat{a}^\dagger_{p\beta}\hat{a}_{q\beta},
    \label{eq:oo_components}
\end{equation}

and the $\kappa_{pq}$ are variational parameters that can be optimized to decrease the energy via orbital rotations. These rotations can be implemented efficiently in a classical computer by producing a new Hamiltonian with updated molecular integrals, and therefore do not imply additional circuit costs.

ADAPT-VQE-SCF optimizes the molecular orbital basis along with the ansatz parameters (i.e., it carries out a simultaneous optimization of $\pmb{\kappa}$ and $\pmb{\theta})$ in each iteration). The gradients of the orbital rotation coefficients $\kappa_{pq}$ can be obtained from the two-particle reduced density matrices, which are measured when evaluating the energy itself; therefore, these gradients are provided to us `for free' during the optimization. However, $\pmb{\kappa}$ comes with an additional $\frac{N_s(N_s-1)}{2}$ variational parameters (with $N_s$ the number of spatial orbitals). This results in a higher dimensional optimization, which might increase the total number of energy and ansatz gradient measurements required for the optimizer to converge and, therefore, indirectly lead to additional measurement costs.

Figure \ref{fig:H6_r=1.5_oo} shows the result of
merging orbital optimization (OO) with the CEO-ADAPT-VQE* algorithm. We can see that the impact on the relevant markers is negligible: the CNOT count and depth required to reach a given error are only very slightly decreased, while the measurement costs are slightly increased. We note that ADAPT-VQE-SCF was developed to enable simulations with large atomic orbital basis sets; it is not surprising that it has a subpar performance in the case of minimal basis sets. 

While we recognize that orbital optimization may be an important enhancement to ADAPT-VQE in the case of larger molecules and/or basis sets, we exclude it from the algorithms in the main text due to its minor impact on the energy under the circumstances we consider.
 
\section{Hamiltonian Grouping}
\label{ap:h_grouping}

\begin{figure*}[htbp]

    \includegraphics{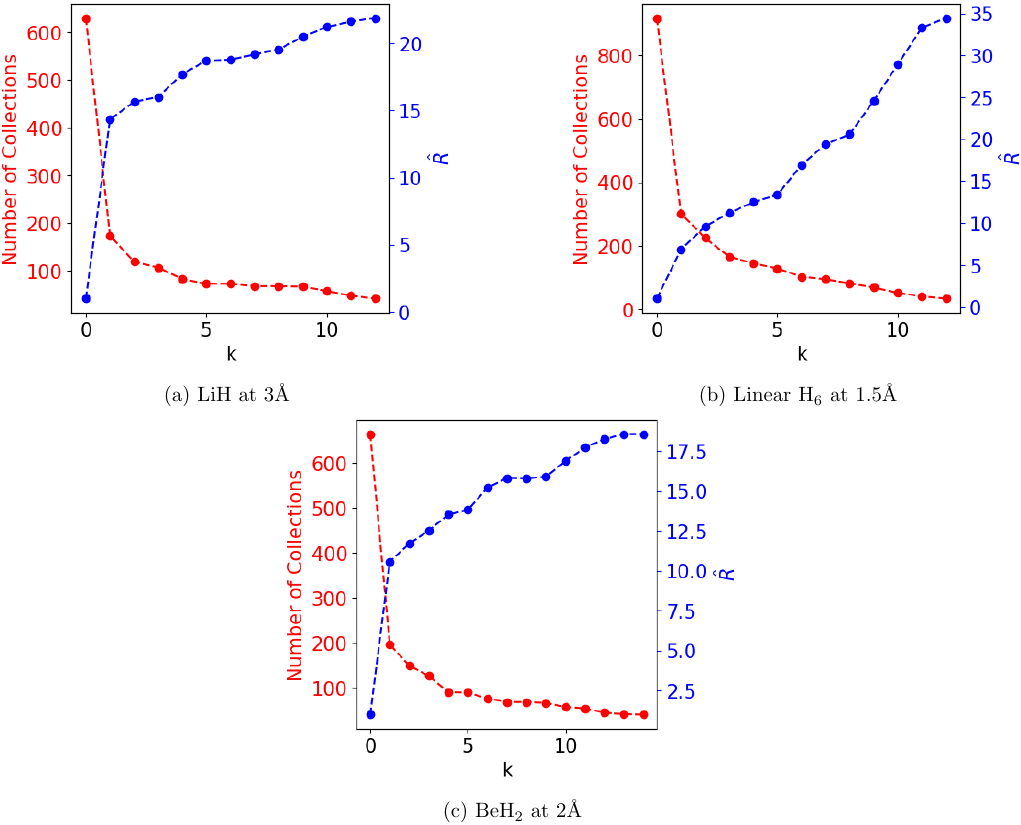}
    \caption{Impact of $k$ on the $k$-Commutativity Grouping}
    \medskip
    \small
    
     Evolution of the number of Pauli string collections and $\hat{R}$ against $k$, which defines the granularity of the commutativity considered in the observable grouping \cite{dalfavero2024}, for \textbf{a} LiH, \textbf{b} H$_6$, \textbf{c} BeH$_2$. While it is expected that fewer collections will be associated with lower measurement costs, covariances and varying coefficient magnitudes result in a nonlinear relation between the two. The metric $\hat{R}$ \cite{Crawford2021} approximates the ratio of measurement costs with and without grouping. The point $k=0$ represents no grouping, such that $\hat{R}$ is one and the number of collections is the total number of Pauli strings in the Hamiltonian.
    
    \label{fig:k_comm}
\end{figure*}

In part of the results included in the main text we applied the $k$-commutativity grouping \cite{dalfavero2024} to the molecular Hamiltonians, and took the $\hat{R}$ metric as an approximation to the ratio $M_u/M_g$ between measurement costs without ($M_u$) and with ($M_g$) grouping \cite{Crawford2021}. As discussed, we chose $k=n$ (general commutativity) in light of the expectation that the depth of the measurement circuit will not be significant relative to the depth of the state preparation circuit. In this case, it is convenient to consider full commutativity to maximize the reduction in measurement costs at the expense of negligible additional circuit depth. However, in general, choosing the optimal $k$ involves a compromise between circuit depth and measurement costs. As such, it is interesting to analyze the evolution of the savings in measurement costs against $k$.

Figure \ref{fig:k_comm} showcases this evolution for the systems we consider in this paper. We observe that for LiH$_2$ and BeH$_2$, the case $k=1$ (qubit-wise commutativity) offers the greatest improvement in measurement costs as compared to other unit increments of $k$. However, in the case of H$_6$, the molecule where the grouping achieves the greatest savings, the improvement resulting from incrementing $k$ is roughly consistent throughout the whole plot.

Unlike Ref.~\cite{dalfavero2024}, we do not observe $\hat{R}$ to stall after a given value $k^*$ - in fact, $\hat{R}$ steadily increases until reaching the maximum value for the highest value of $k$. We conjecture that this difference may stem from the following facts: (i) the systems we consider are distinct from those where such behavior was encountered (Fermi-Hubbard vs molecular Hamiltonians), and (ii) the number of qubits of our systems is significantly lower. The stabilization of $\hat{R}$ at a maximum value for $k<n$ could happen for larger $n$. In that case, the measurement cost reduction of general commutativity can be achieved with shallower measurement circuits.

\end{document}